\documentclass[aps,amsmath,amssymb,superscriptaddress,onecolumn,floatfix]{revtex4}
\usepackage{amssymb,amsmath,bm,hyperref}

\usepackage{mathrsfs}
\usepackage{cases}
\usepackage{graphicx}
\usepackage{subfigure}
\usepackage{here}

\usepackage[normalem]{ulem}

\usepackage{xcolor}

\makeatletter
\def\be{\begin{equation}}
\def\ee{\end{equation}}
\def\bea{\begin{eqnarray}}
\def\eea{\end{eqnarray}}
\def\mc{\mathcal}

\begin{document}

\title{Magnetic excitations in the one-dimensional Heisenberg-Ising model with external fields and their experimental realizations}
\date{\today}
\author{Jiahao Yang}
\affiliation{Tsung-Dao Lee Institute, Shanghai Jiao Tong University, Shanghai 200240, China}

\author{Xiao Wang}
\affiliation{Tsung-Dao Lee Institute, Shanghai Jiao Tong University, Shanghai 200240, China}

\author{Jianda Wu}
\email{wujd@sjtu.edu.cn}
\affiliation{Tsung-Dao Lee Institute, Shanghai Jiao Tong University, Shanghai 200240, China}
\affiliation{School of Physics and Astronomy, Shanghai Jiao Tong University, Shanghai 200240, China}

\begin{abstract}
The one dimensional (1D) spin-1/2 Heisenberg-Ising model, a prototype quantum many-body system,
has been intensively studied for many years.
In this review, after a short introduction on some basic concepts of group theory for the octahedral group,
a detailed pedagogical framework is laid down to derive the low-energy effective Hamiltonian
for the Co-based materials. The 1D spin-1/2 Heisenberg-Ising model is obtained
when applying the analysis to quasi-1D antiferromagnetic materials
$\rm BaCo_2V_2O_8$ and $\rm SrCo_2V_2O_8$.
After the preparation, we review the theoretical progresses of a variety of
novel magnetic excitations and
emergent physics in the 1D spin-1/2 Heisenberg-Ising model,
and further summarize their recent experimental realizations.

\end{abstract}

% % \noindent{\it Keywords}:
% 1D spin-1/2 Heisenberg-Ising model,
% low-energy effective Hamiltonian,
% octahedral  group,
% crystal field potential,
% quasi-1D antifferomagnetic materials,
% spinon, string, (anti)psinon,
% transverse field Ising chain universality,
% quantum $E_8$ integrable model,
% $E_8$ particles

% \submitto{J. Phys.\ A: Math.\ Theor.}

\maketitle
%==============================================================

%%%%%%%%%%%%%%%%%%%%%%%%%%%%%%%%%%%%%%%%%%
\section{Introduction}

The one-dimensional (1D) Heisenberg model and its extension
1D Heisenberg-XY/Ising model were born many decades ago \cite{bethe1931,Kasteleijn_1952104,Sutherland_1970}.
Due to their quantum integrability and algebraic beauty \cite{Jimbo1995,Faddeev1979,slavnov_algebraic_2019,Baxter_1982_ExactlySM,caux_remarks_2011},
close-entwined physical and mathematical progress have been continuously surging since the birth of the models.
The studies not only significantly deepen and extend our understanding
on magnetic excitations and many-body physics,
but also expand frontiers of mathematics and even
create new mathematical directions.
From physical side, the studies reveal a series of charming
many-body excitations and rich emergent phenomena.
The revealed excitations include fractional
types such as spinon \cite{FADDEEV_spinwave_1981,faddeev_spectrum_1984,Muller_spinon_1981,Karbach_spinon_1997,Bougourzi_spinon_1996,Bougourzi_spinon_1998,caux2008}, (anti)psinon \cite{Karbach_pp_pap_2002,karbach2000III},
and exotic ones such as string excitations \cite{bethe1931,takahashi_1D_1971,Gaudin_XXZ_1971,Taka_suzuki_XXZ_1972,Takahashi1999},
where each of the former two excitations always appears in pairs while the later one always
contains bounded magnons.
Depending on the way to tune the anisotropy of the model and how the magnetic field is applied,
a variety of emergences appear:
Tomonaga-Luttinger liquid (without field
or with longitudinal field)
\cite{tomonaga_remarks_1950,Luttinger_1960,Luther_TLL_1975},
integrability with supersymmery (with a fine-tuned anisotropy without field) \cite{Fendley_SUSY_2003},
transverse-field Ising universality (with transverse field) \cite{Pfeuty1970,sachdev_2011,Jianda_crossover_2018}, and $E_8$
physics (with transverse and longitudinal fields)
\cite{a_b_zamolodchikov_integrals_1989,jianda_E8_2014,DELFINO_1995,xiao_cascade_2021,DELFINO199440}.
From mathematical side,
the Yang-Baxter equation and algebraic structure discovered in the model
significantly extend
our understandings on algebraic structure, and play an essential role
in building up the mathematical framework for quantum group and quantum algebra
\cite{Jimbo1995,Faddeev1979,slavnov_algebraic_2019,Baxter_1982_ExactlySM,drinfeld_hopf_1990,drinfeld_quantum_1988,drinfeld_Yangians_1988,drinfeld_QHA_1990,drinfeld_YBE_1983,Maillet_Drifeld_twists_2000}.

Along the course, experimental probes of the aforementioned exciting physics meet
great difficulties in early years due to limited techniques and lack of clean crystals.
With the experimental developments, the first progress is made for
the realization of spinon in the $\rm SrCuO_2$ \cite{kim_spinon_2006,Zaliznyak_spinon_2004}
and even higher-order spinon states are observed in $\rm Sr_2CuO_3$ \cite{mourigal_XXX_2013}.
Soon after the first observation of the
spinon, the magnetic excitations with clear Tomonaga-Luttinger liquid behavior
are also observed in the $\rm BaCo_2V_2O_8$ (BCVO) \cite{Kimura_TLL_2007}.
However, the demonstrations in material for the string excitations,
transverse field Ising chain (TFIC) universality along with the exotic $E_8$ physics
remain as a challenge until recently. The difficulty is largely due to
the challenge to figure out proper parameter regions for the
real material to realize those exotic physics.
Substantial progresses are made with a series of theoretical work
\cite{Jianda_crossover_2018,jianda_E8_2014,xiao_cascade_2021,Zou_universality_2019,yang_string_one-dimensional_2019} which
clearly identify proper parameters to realize those physics in corresponding
materials.
Meanwhile, experiments are carried out and unambiguously realize for
the first time the long-desired physics in corresponding quasi-1D materials
\cite{kimura_high_2006,bera_magnetic_2014,wang_string_experimental_2018,bera_string_dispersions_2020,cui_tfic_quantum_2019,wang_tfic_quantum_2018,zou_e_8_2021,zhang_e8_observation_2020,amelin_e8_experimental_2020,yang_Local_2022}.

\begin{figure}[h]
\centering
\includegraphics[width=0.6\textwidth]{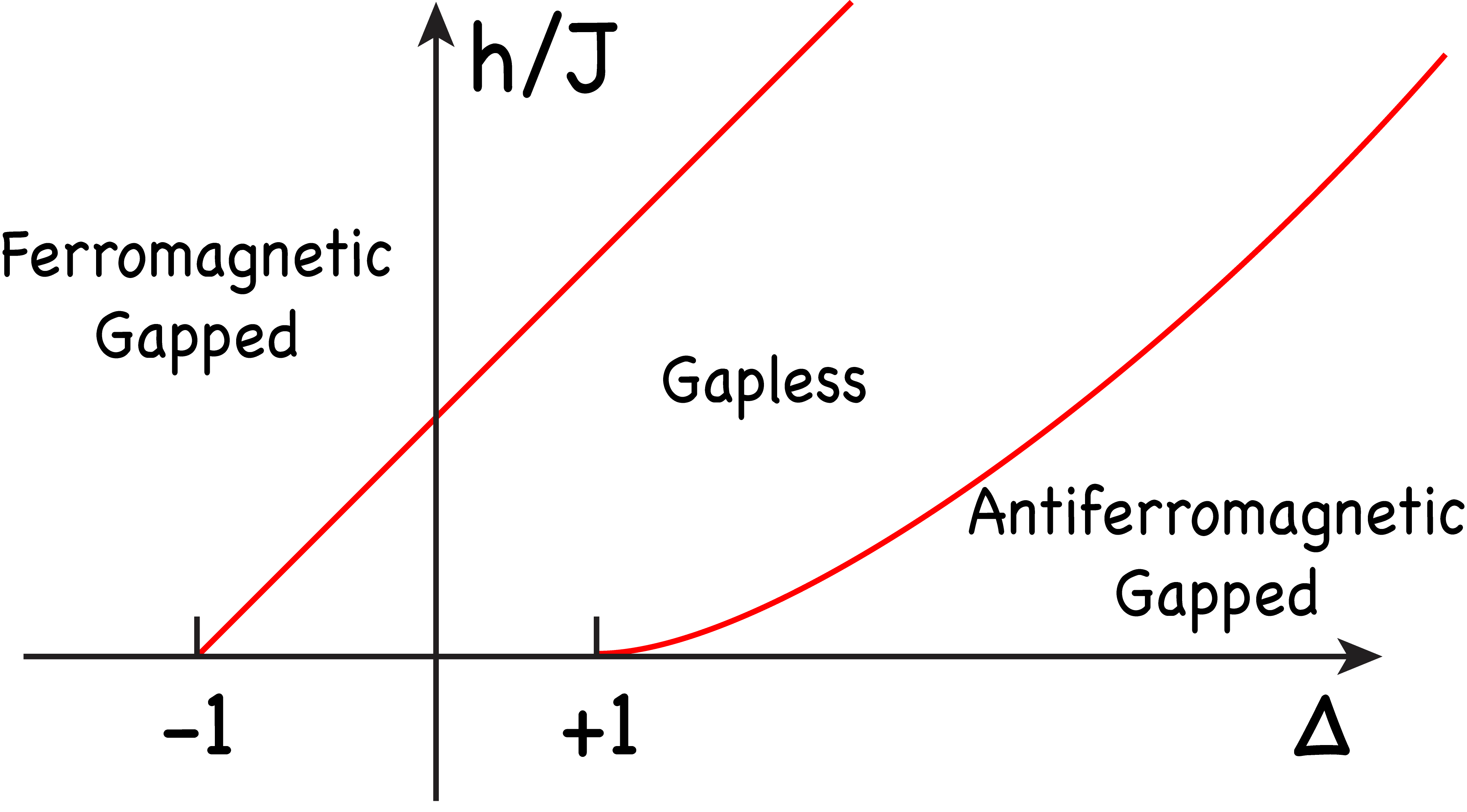}
\caption{
Phase diagram of the XXZ model Eq.~\eqref{eq:H_XXZ_lh}.
}
\label{fig:phase_XXZ}
\end{figure}

In this review, we focus on the 1D spin-1/2 Heisenberg-Ising model (XXZ model) with Ising anisotropy $\Delta>1$,
\be
H=J\sum_{i=1}^N \left(S^x_iS^x_{i+1}+S^y_iS^y_{i+1}+\Delta S^z_iS^z_{i+1}
\right)
-h\sum_{i=1}^N S^z_i,
\label{eq:H_XXZ_lh}
\ee
with the spin component  $S_i^\alpha = \sigma_i^\alpha/2$ at site $i$ $(\alpha=x,y,z; \;{\rm and}\; \sigma_i^\alpha
\;{\rm labels\;the\; Pauli\; matrix})$,
the coupling $J>0$ between neighbouring spins,
and the external field $h$ along $z$ direction.
The phase diagram for the model is illustrated in Fig.~\ref{fig:phase_XXZ} \cite{CNYang_XXZ_1966,Johnson1991,Franchini2017}.
Since total spin along $z$-direction $S^z_{T}$ is conserved,
the Hilbert space can be divided into different subspaces according to
the quantum numbers of $S^z_{T}$ which is the magnetization along $z$ direction.
If $\Delta=1$, the Hamiltonian recovers the XXX model with an $SU(2)$ symmetry.

\begin{figure}[h]
\centering
\includegraphics[width=0.6\textwidth]{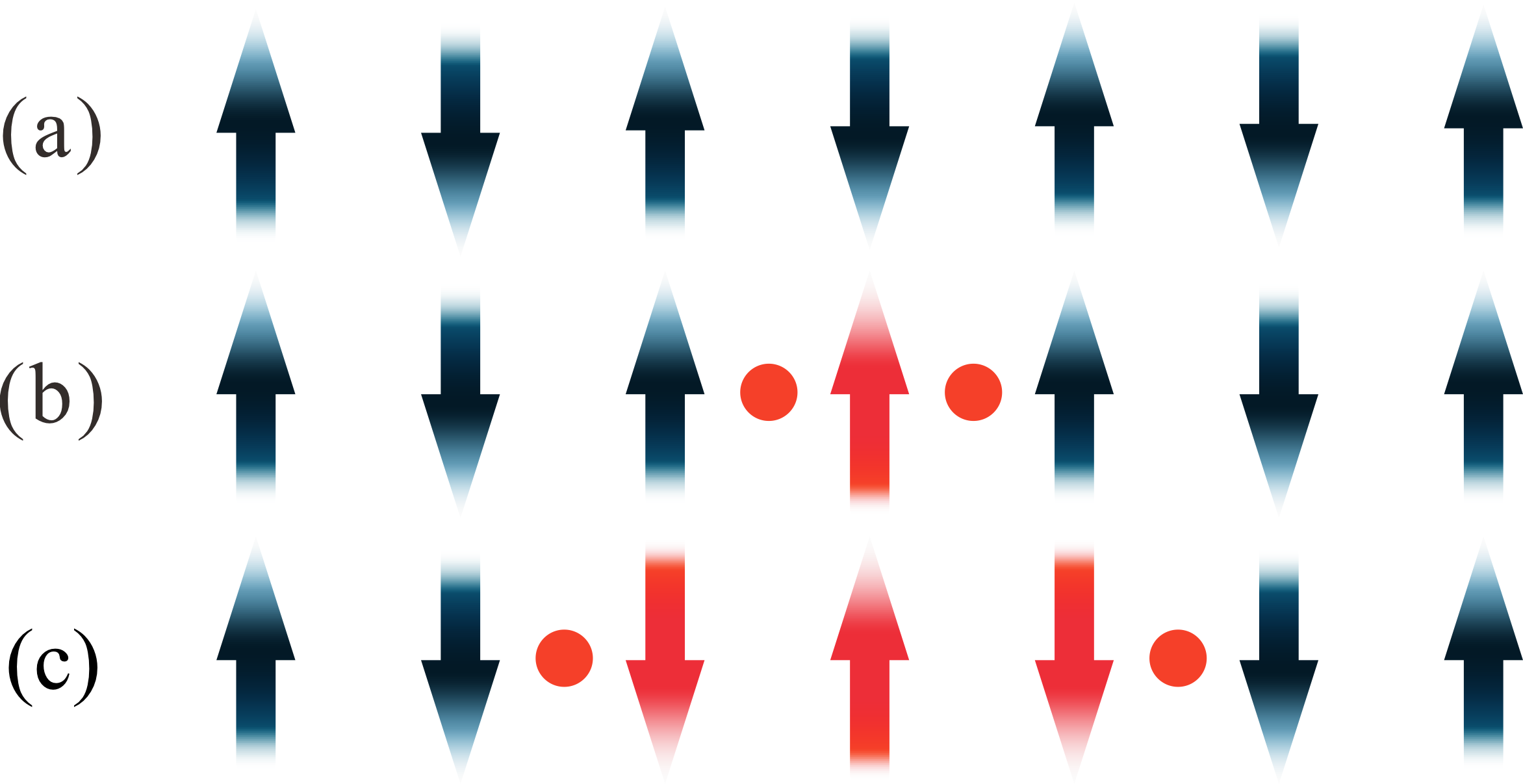}
\caption{From the AFM ground state (a), a pair of spinons illustrated as two domain walls,
created by a single spin-flip (b), propagate along the chain (c). }
\label{fig:spinon}
\end{figure}

At zero field $h=0$, the 1D spin-1/2 XXZ model, has an antiferromagnetic (AFM) ground state.
Its elementary excitations referred to spinons are fractional spin-1/2 quasiparticles
appear in pairs in the spin dynamics.
The  dynamical structure factor of two-spinon excitations can be calculated exactly
in both transverse and longitudinal channels \cite{caux2008,castillo_exact_2020}.
The two-spinon spectrum is directly observed in $\rm SrCo_2V_2O_8$ (SCVO) by inelastic neutron scattering (INS)
and agrees with theoretical prediction \cite{bera_spinon_2017}.

\begin{figure}[h]
\centering
\includegraphics[width=0.5\textwidth]{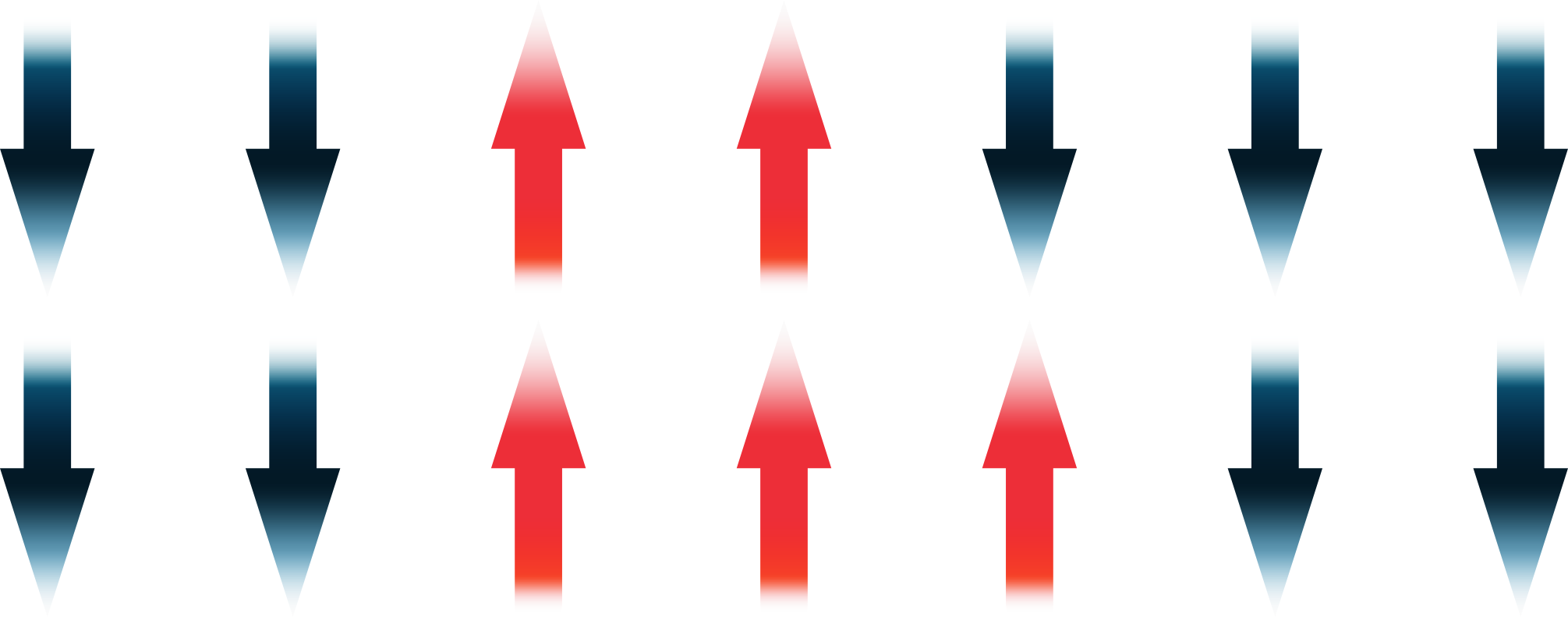}
\caption{An illustration for the 2 (3)-string excitation [red arrows in top (bottom) row] in spin configurations. }
\label{fig:2string}
\end{figure}

One appealing excitation in Heisenberg models is the string state which is a magnetic soliton state \cite{bethe1931,takahashi_1D_1971,Gaudin_XXZ_1971,Taka_suzuki_XXZ_1972}.
In the Ising limit ($\Delta \gg 1$) the magnetic soliton corresponds to
a bound object of flipped spins [Fig.~\ref{fig:2string}].
Despite of long-time theoretical recognition and continuously
theoretical studies of the
string excitations and their essentially compelling many-body nature \cite{bethe1931,Takahashi1999},
%a realization of these exotic states
%places a great challenge for experiments and
feasible proposals to
directly observe those exotic states remain a challenge \cite{ishimura_dynamical_1980,Imambekov_1D_2012,kohno_string_dynamically_2009,Pereira_edge_2008,Pereira_spectral_2009,Caux_comput_2005,Caux_computation_2005,Shashi_Nonuniversal_2011, Ganahl_Observation_2012}.
The difficulty lies in locating a proper parameter region
such that the string contribution to the spin dynamics is non-negligible.
A silver lining appears in 2017 that
the Ref.~\cite{yang_string_one-dimensional_2019}
proposes a promising
field-induced quantum critical region of the 1D spin-1/2 XXZ model Eq.~\eqref{eq:H_XXZ_lh},
to directly probe the string excitations, where
the string excitations are shown to dominate the dynamic spectrum in the
quantum critical region with small magnetization.
Following the proposal, the
string excitations are clearly detected and observed
in the zone center of SCVO for the first time
via the high-resolution terahertz (THz) spectroscopy measurement \cite{wang_string_experimental_2018}.
Under concrete theoretical guidance, the exotic 2- and 3-string excitations
as well as novel low energy fractional
magnetic excitations are identified in the field-induced quantum-critical region.
In 2020, with INS measurments on the same material,
complete dispersions of the string exciations
over the full Brillouin zone are obtained \cite{bera_string_dispersions_2020}.
In both experiments, the obtained excitation spectrum
and its magnetic field dependency
perfectly agree with theoretical calculations from low to high energy,
demonstrating a rare success in understanding strongly correlated magnetic systems.

When the longitudinal field is replaced by a transverse field
in Eq.~\eqref{eq:H_XXZ_lh}, a quantum phase transition arises with
field tuning. The corresponding quantum criticality around the
quantum critical point (QCP) falls into the class of
TFIC universality \cite{dmitriev_1D_2002}.
%near the field-tuned quantum critical point (QCP)
Near the QCP of the TFIC universality,
the Gr\"uneisen ratio,
the ratio of magnetic expansion coefficient to specific heat \cite{Yu_Gruneisen_2020},
is found to exhibit a unique singular behavior:
it can be either divergent or convergent when the system approaches
its QCP with field or
temperature tuning, respectively \cite{Jianda_crossover_2018}.
The exotic singular behavior
can serve as a smoking gun to
justify the TFIC universality in real materials.
Meanwhile, a relevant experiment is carried out for BCVO
in the transverse field \cite{wang_tfic_quantum_2018}, where
a 1D QCP appears at a strong field.
Near the 1D QCP, the obtained quantum critical behaviors
for the Gr\"uneisen ratio exactly
follow the aforementioned scaling relation in the class of TFIC universality,
which thus beautifully realizes the TFIC universality.

More surprisingly, when the quantum critical TFIC is perturbed
by a longitudinal field, an integrable massive relativistic quantum field theory,
dubbed as quantum $E_8$ integrable model, further emerges. In the model, it
contains eight types of massive particles whose scattering can be fully described
by the $E_8$ exceptional Lie algebra \cite{a_b_zamolodchikov_integrals_1989}.
In 2010, an INS experiment on the quasi-1D ferromagnetic
material $\rm CoNb_2O_6$ (CNO) observes
two lowest excitations whose energy ratio matches the mass ratio of the two lightest $E_8$
particles, providing a preliminary evidence for the existence of the exotic $E_8$ physics
in real material \cite{coldea_quantum_2010}.
However, the measured DSF in the continuum region is not accurate enough
to distinguish other $E_8$ particles,
which makes inconclusive a complete realization of the $E_8$
physics in the material CNO.
In 2020, a further THz spectroscopy experiment is performed on the same material,
which obtained a detailed spectrum in the continuum region
\cite{amelin_e8_experimental_2020}.
However, the obtained spectrum in the continuum region apparently deviates
from theoretically results \cite{xiao_cascade_2021},
implying more systematic and careful studies are needed
to conclude
the realization of the $E_{8}$ physics in the CNO \cite{morris_duality_2021,fava_glide_2020}.
In the same year,
by taking advantage of the delicate microscopic structure of BCVO,
combined efforts made from theoretical analysis,
numerical simulation,
THz spectroscopy, nuclear magnetic resonance (NMR), and INS experiments unambiguously reveal
beautiful $E_8$ physics for the first time in the material,
which not only confirms the eight single $E_8$
particles but also the excitations composed by multi-$E_8$ particles.
\cite{xiao_cascade_2021,zou_e_8_2021,zhang_e8_observation_2020}.

Above brief summary clearly shows rich physics in the Co-based materials, such as BCVO, SCVO, and CNO etc.,
where the $\rm Co^{2+}$ ion carries on the magnetism with $3d^7$ electrsons.
The ground state of the local $\rm Co^{2+}$ ion possesses
total orbital angular momentum $L=3$ and total spin $S_{tot} = 3/2$ \cite{Fazekas1999}.
Due to the superexchange mechanism the interaction of the neighbouring spin is isotropic, thus the effective model
becomes a spin-3/2 XXX model \cite{Anderson1959,Anderson1994,SAWATZKY1976}.
In those materials, the local environment and the spin-orbit coupling (SOC) are non-negligible,
which can cause further splitting of the local ground state manifold.
Since time-reversal symmetry is still preserved, the new ground state
in general at least have
a degeneracy of Kramers doublet. If we project the spin-3/2 XXX model to
the new ground state, we can obtain an effective 1D spin-1/2 XXZ model to
describe the low-energy physics, as manifested in
a large number of Co-based materials
\cite{shiba_exchange_2003,lines_magnetic_1963,Sarte2018}.

The remainder of the review is organized as follows.
Sec.~\ref{sec:Ogroup} introduces some basic properties of octahedral group and atomic basis for the $\rm CoO_6$ octahedron.
Sec.~\ref{sec:Hamiltonian} is devoted to derive an effective 1D spin-1/2 XXZ model for the Co-based materials whose ground state is a Kramers doublet, and gives a detailed discussion for the Zeeman effect in SCVO and BCVO.
Sec.~\ref{sec:excitations} reviews recent theoretical progresses on the magnetic excitations and emergent physics in the 1D spin-1/2 XXZ model without or with the presence various magnetic fields, and summarizes their experimental realizations.
Sec.~\ref{sec:conclusion} is the conclusions.

\section{The Octahedral Group and Atomic Basis}
\label{sec:Ogroup}

\begin{figure}[ht]
\centering
\includegraphics[scale=0.7]{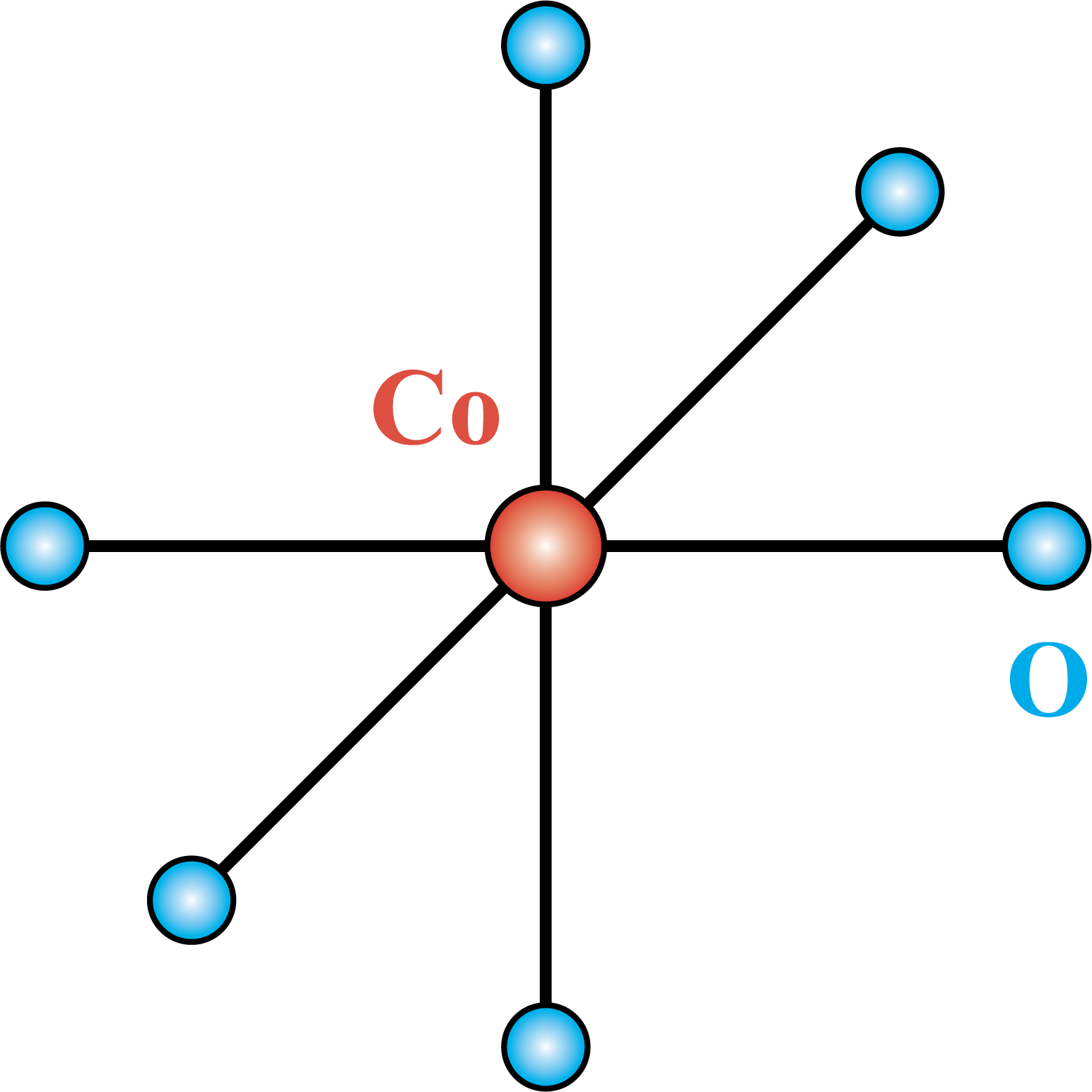}

\caption{The local octahedron environment for a $\rm Co^{2+}$ with six oxygen ions surrounded.
}
\label{fig:Co_octa}
\end{figure}
\begin{figure}[ht]
\centering
\includegraphics[scale=0.4]{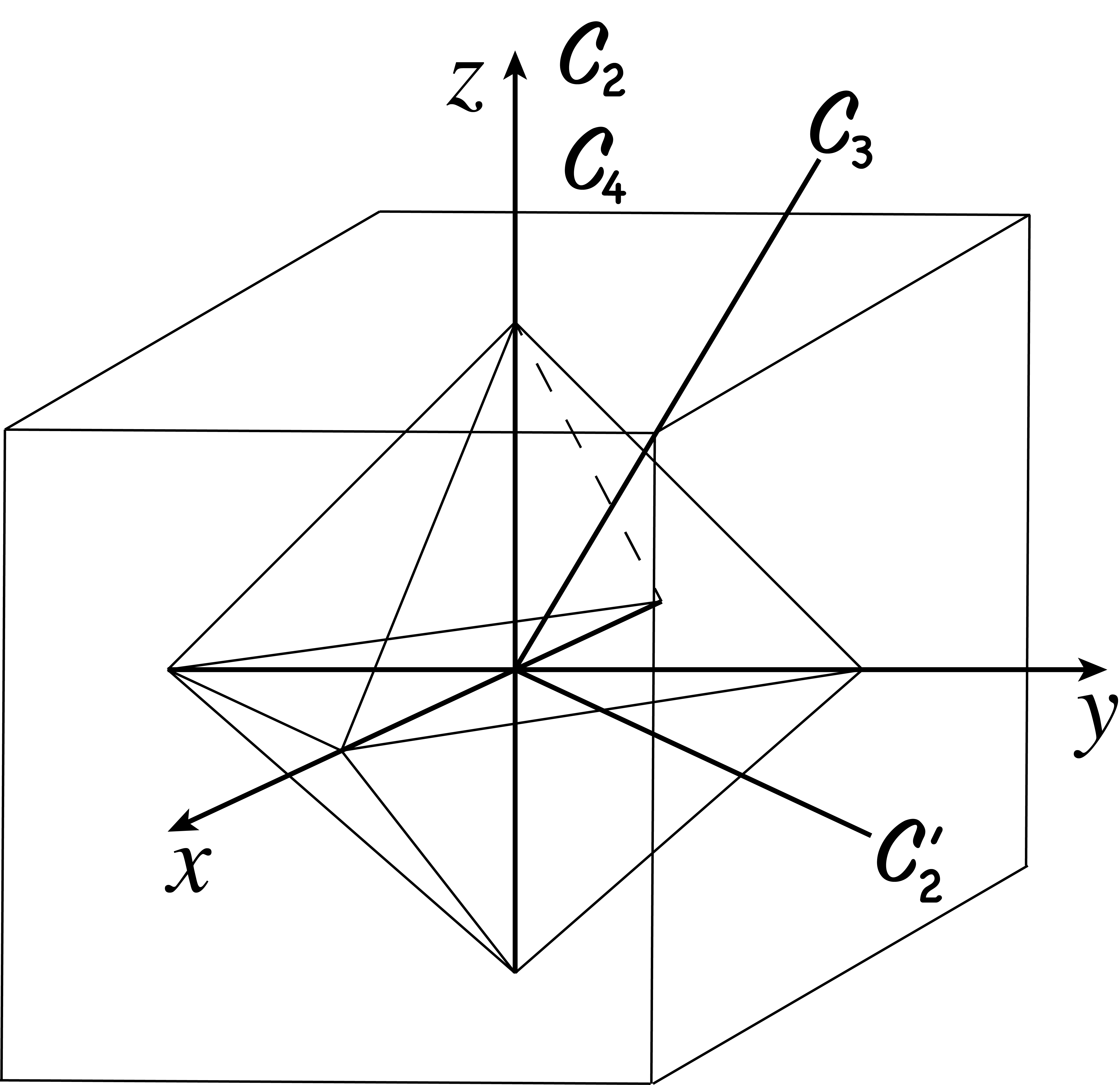}
\caption{Equivalent axes belonging to $\mathcal{O}$ symmetry of the octahedron and the cube.}
\label{fig:Ogroup}
\end{figure}

For the local octahedron environment
the crystal field produced by six oxygen ions in the octahedron $\rm CoO_6$ has the octahedral symmetry described by $\mathcal{O}$ group,
consisting of all the rotations that keep an octahedron invariant [Fig.~\ref{fig:Co_octa}].
$\mathcal{O}$ group can also characterize the symmetry of
a crystal field induced by eight corner atoms of a cube [Fig.~\ref{fig:Ogroup}].
Thus, the crystal field of an octahedron is also referred to the ``cubic field''.

From Fig.~\ref{fig:Ogroup}, we can enumerate the elements of $\mathcal{O}$ group,
and distribute them to different classes
of inequivalent sets
of operations \cite{Dresselhaus2008,Zee2016,Tinkham1964}.
Rotations by the same angle about equivalent axes form a class
and two axes are equivalent if they are related to each other by a symmetry operation \cite{Fazekas1999}.
Thus, there are six $\pi/2$- and $3\pi/2$-rotations in the $\mathcal{C}_4$ class.
The $\pi$-rotations about the coordinate axis form the three-element class $\mathcal{C}_2$.
And there is another type of $\pi$-rotation $\mc{C}_2'$ about axes, where the axes pass through
the centre of the cube and are parallel to the face diagonals.
Rotations by $\pi/3$ and $2\pi/3$ about body diagonals of the cube generate the eight-element class $\mc{C}_3$.
Finally, the identity forms a class in itself.
In total, there are five different classes in $\mathcal{O}$ group with 24 group elements.

In group theory \cite{Dresselhaus2008,Zee2016}, the action of a symmetry operation $\hat R$ on a given
basis set can be described by a matrix $M({R})$.
The matrices of all symmetry operations form a group which is homomorphic to the corresponding symmetry group.
And the matrix group is referred to a
representation of the corresponding symmetry group.
The representation characterizes the symmetry properties which must be independent of the choice of the basis.
Furthermore,
the trace of a matrix representation is invariant under basis transformation,
which is introduced to serve as a character of the representation,
\be
\chi(R)=\sum_i M_{ii}(R)=\text{Tr}(M(R)),
\ee
For the identity $E$ of the group, $\chi(E)=d$, the
dimensionality of the representation.
For the $\mc{O}$ group, Table \ref{tab:cha_O} lists the characters of
all classes and irreducible representations.
The 1D representation $A_1$ is known as the identity representation,
and other four irreducible representations have dimensions 1, 2, 3, 3.

\begin{table}[t!]
\centering
\begin{center}
\begin{tabular}{c|c|c|c|c|c|c|c}
    &  &basis  & $ E $ & $8\mathcal{C}_3$ & $3\mathcal{C}_2$ & $6\mathcal{C}_2'$ & $6\mathcal{C}_4$ \\
\hline
$A_1$ 	  & $ \Gamma_1 $ & $\{x^2+y^2+z^2\}$     & 1 & 1  & 1  & 1  & 1  \\
$A_2$    & $ \Gamma_2 $ & $\{xyz\}$              & 1 & 1  & 1  & -1 & -1  \\
$E$     & $\Gamma_3$   & $\{x^2-y^2,3z^2-r^2\}$ & 2 & -1 & 2  & 0  & 0  \\
$T_1$ & $\Gamma_4$   & $\{x,y,z\}$            & 3 & 0  & -1 & -1 & 1  \\
$T_2$ & $\Gamma_5$   & $\{xy,yz,zx\}$         & 3 & 0  & -1 & 1  & -1
\end{tabular}
\end{center}
\caption{ The character table of the $\mathcal{O}$ group with two standard notations of the irreducible representations.
One commonly used basis set is also listed. }
\label{tab:cha_O}
\end{table}

A representation $\Gamma$ of a group is reducible if we can find a transformation $U$ which makes all matrices block-diagonal.
Then the reducible representation $\Gamma$ can be decomposed
into a direct sum of irreducible representations $\Gamma_j$,
\be
\Gamma=\bigoplus_{j}a_j \Gamma_j.
\label{eq:decompose}
\ee
where $a_j$ is the number of times that the irreducible representations occur.
Using the characters of different classes,  $a_j$  can be determined as follows,
\be
a_j=\frac{1}{\mathcal{N}}
\sum_{k=1}^r
\mathcal{N}_k \chi_j^*(\mathcal{C}_k)\chi(\mathcal{C}_k).
\ee
where the summation is over the classes and $\mathcal{N}_k$ is the number of elements in the $k$th class $\mathcal{C}_k$ with $\sum_k\mathcal{N}_k=\mathcal{N}$, the total number of the group elements.
And the $\chi(\mathcal{C}_k)$ and $\chi_j(\mathcal{C}_k)$ are the characters of the
reducible representation $\Gamma$ and irreducible representation $\Gamma_j$
for the class $\mathcal{C}_k$, respectively.
\begin{table}[ht]
\centering
\begin{center}
\begin{tabular}{c|c|c|c|c|c}
    &  $ E $ & $8\mathcal{C}_3$ & $3\mathcal{C}_2$ & $6\mathcal{C}_2'$ & $6\mathcal{C}_4$
    \\\hline
$\Gamma_{l=0}$  & 1 & 1 & 1 & 1 & 1    \\
$\Gamma_{l=1}$  & 3 & 0 & -1 & -1 & 1    \\
$\Gamma_{l=2}$  & 5 & -1 & 1 & 1 & -1    \\
$\Gamma_{l=3}$  & 7 & 1  & -1 & -1 & -1
\end{tabular}
\end{center}
\caption{A character table of the representations of the octahedral group
$\mathcal{O}$ with different orbital angular momentum $l = 0,1,2,3$.}
\label{tab:cha_spdf}
\end{table}

At the level of the atomic (ionic) Hamiltonian, the angular dependence of the atomic wave functions is described by the familiar spherical harmonics $Y_l^m(\theta,\varphi)$.
This comes from the fact that the pure atomic potential is spherically
symmetrical and the symmetry group of the Hamiltonian is the continuous group $SO(3)$
which contains all rotations in three-dimensional space.
As a result, the (total) angular momentum $l$ is a good quantum number
with eigenstates classified by $l$.
For an isotropic system, all rotation axes are equivalent thus
rotations around equivalent axes by a same angle $\alpha$ belong to a same class.
The corresponding character of an $\alpha$-rotation is
$\chi^l(\alpha)={\sin[(l+1/2)\alpha]}/{\sin(\alpha/2)}$ where $l$ is the angular momentum quantum number.
Since the octahedral group is a subgroup of the rotation
group $\mathcal{O}\in SO(3)$, every irreducible
representation of $SO(3)$ can provide a representation of $\mathcal{O}$, which usually is reducible.
In Table \ref{tab:cha_spdf}, we list the characters of four representations with different orbital angular momentum $l$.
As for the $\rm Co^{2+}$ with $L=3$, the seven-fold representation $\Gamma_{L=3}$ is reducible in $\mathcal{O}$ symmetry.
Following Eq.~\eqref{eq:decompose},
$\Gamma_{L=3}$ can be decomposed into a direct
sum of a singlet $A_2$ and  two triplets $T_1$ and $T_2$, i.e.,
$\Gamma_{L=3}=A_2\oplus T_1\oplus T_2$.
Within this seven dimensional Hilbert space, we are free to choose
any seven linearly independent basis which follow the transformation of $\mathcal{O}$ group as a basis set.
These seven basis functions can be constructed after taking into account the cubic symmetry \cite{AABB1970,Bleaney_1953}:
\be
\small
% T_1\ \mbox{manifold}\
\begin{cases}
|T_1,2\rangle=\sqrt{\frac{3}{8}}Y_3^{-1}
+\sqrt{\frac{5}{8}}Y_3^3\\
|T_1, 0\rangle=Y_3^0\\
|T_1,-2\rangle=\sqrt{\frac{3}{8}}Y_3^{1}
+\sqrt{\frac{5}{8}}Y_3^{-3}
\end{cases},
% T_2\
\begin{cases}
|T_2,2\rangle=\sqrt{\frac{5}{8}}Y_3^{-1}
-\sqrt{\frac{3}{8}}Y_3^3\\
|T_2,0\rangle=\frac{1}{\sqrt{2}}(Y_3^2+Y_3^{-2})\\
|T_2,-2\rangle=\sqrt{\frac{5}{8}}Y_3^{1}
-\sqrt{\frac{3}{8}}Y_3^{-3}
\end{cases},
|A_2\rangle=\frac{1}{\sqrt{2}}(Y_3^2-Y_3^{-2}).
\label{eq:basis}
\ee
It's easy to verify that the above basis functions are  invariant under rotation within corresponding subspace and the trace of matrix representation of every rotation operation is consistent with the character table \ref{tab:cha_O}.

Group theory alone can not distinguish which irreducible representation is the lowest lying one.
To figure out this, the details of the Hamiltonian have to be specified.
{So next we shall find out the ground state manifold of $\rm CoO_6$ with the help of crystal field Hamiltonian.}

\section{Crystal Field and Effective Hamiltonian }
\label{sec:Hamiltonian}

In this section,
we show details to determine the lowest-lying Kramers doublet followed by
the crystal field Hamiltonian of the octahedron $\rm CoO_6$ via the
equivalent operator method \cite{AABB1970,Bleaney_1953,hutchings1964}.
Within the obtained ground-state doublet,
an effective XXZ model with pseudospin $S=1/2$ is built up for SCVO and BCVO.

\subsection{Crystal field Hamiltonian}

We start from the point charge model where the electrostatic potential
is generated from the surrounding anions, like the oxygen ions in Fig.~\ref{fig:Co_octa}.
This potential is referred to  crystal field potential (in coordinate representation):
\be
\mathcal{V} ( \mathbf{r} ) = \sum _ { j } \frac { q _ { j } } { \left| \mathbf { R } _ { j } - \mathbf { r }  \right| }
\ee
where $q_j=q=-Ze$ is the charge at the $j$th anion with a distance $|\mathbf { R } _ { j }| = a$ from the origin (e.g., Co$^{2+}$) and $\bf r$ is an arbitrary spatial point from the origin.
Since the crystal field potential $\mathcal{V}$ satisfies Laplace's equation $\Delta \mathcal{V}=0$,
it can be conveniently expanded as a sum of spherical harmonics $Y_n^m(\theta,\varphi)$:
\be
\mathcal{V} ( r , \theta , \varphi ) =
\sum_{n=0}^{\infty}\mathcal{V}_n
=
\sum_{n=0}^\infty \sum_{m=-n}^n A_n^m Y_n^m(\theta,\varphi)r^n
\label{eq:Vexpansion}
\ee
where $A_n^m$ are coefficients to be determined.
Thus, the crystal field potential energy of a magnetic ion at the origin, e.g., Co$^{2+}$ in Fig.~\ref{fig:Co_octa}, is
$W=\sum_i(-e)\mathcal{V}({\bf r}_i)$ where ${\bf r}_i$ is the position of an electron of the cobalt ion.
We need to calculate the matrix element of it between two basis functions, i.e. $\int{\rm d}{\bf r}\phi^*\mathcal{V}\psi$.
In order to calculate it efficiently, we can reduce the right hand side of Eq.~\eqref{eq:Vexpansion} by symmetry argument, since not all the terms in the sum have the same symmetry of surrounding anions.
For instance, all terms with $n>2L$ vanish,
because both $\psi$ and $\phi$ are $L=3$ basis functions Eq.~\eqref{eq:basis} for $\rm Co^{2+}$ ion.
Thus, the terms with $n>6$ vanish due to the orthogonality of spherical harmonics.
By a similar argument,
all terms with odd $n$
vanish
%have zero matrix elements
because $\psi$ and $\phi$ have the same parity, i.e. $\phi^*\psi$ has even parity.
Finally, due to the $C_4$ rotational symmetry along the $z$ direction of the octahedron, the potential $\mathcal{V}(r,\theta,\varphi)$ must be the same as  $\mathcal{V}(r,\theta,\varphi+\pi/2)$.
And as the only $\varphi$ dependence is $\exp(im\varphi)$, this means the only non-vanishing $A_n^m$ are those terms with $m=0$ and $\pm4$.
After neglecting the $Y_0^0$ term, which is a constant, the remaining non-trivial contributions are $\mathcal{V}_4$ and $\mathcal{V}_6$ \cite{hutchings1964}:
\be
\mathcal{V}(\mathbf{r})=\mathcal{V}_4(\mathbf{r})+\mathcal{V}_6(\mathbf{r}) \,,
\label{eq:V46}
\ee
where
\be
\begin{split}
\mathcal { V } _ { 4 }(\mathbf{r})
&=
D_4
\left[
Y_4^0(\theta,\varphi)
+\sqrt{\frac{5}{14}}
\left(
Y_4^4(\theta,\varphi)
+
Y_4^{-4}(\theta,\varphi)
\right)
\right]r^4
\\&=
A_4 \left( x ^ { 4 } + y ^ { 4 } + z ^ { 4 } - \frac { 3 } { 5 } r ^ { 4 } \right)
\end{split}\,,
\label{eq:V4}
\ee
and
\be
\begin{split}
\mathcal { V } _ { 6 } (\mathbf{r})
&=
D_6 \left[
Y_6^0(\theta,\varphi)
-\sqrt{\frac{7}{2}}
\left(
Y_6^4(\theta,\varphi)
+Y_6^{-4}(\theta,\varphi)
\right)
\right]r^6
\\&=
A_6 \left[ x ^ { 6 } + y ^ { 6 } + z ^ { 6 } + \frac { 15 } { 4 } \left( x ^ { 2 } y ^ { 4 } + x ^ { 2 } z ^ { 4 }
+ y ^ { 2 } x ^ { 4 } + y ^ { 2 } z ^ { 4 } + z ^ { 2 } x ^ { 4 } + z ^ { 2 } y ^ { 4 } \right) - \frac { 15 } { 14 } r ^ { 6 } \right]
\end{split}\,.
\label{eq:V6}
\ee
The prefactors depend on the detailed geometry of the anions.
For instance, if the potential $\mathcal{V}(x,y,z)$ is generated by anions $q$ at the corners of an octahedron;
i.e., at $(a,0,0)$, $(0,a,0)$, $(0,-a,0)$, $(0,0,a)$, $(0,0,-a)$, one finds [{\it c.f.} appendix.~\ref{app:Oct_V}]
\be
D_4=\frac{7\sqrt\pi}{3}
\frac{q}{a^5}
,\quad
D_6=\frac{3}{2}
\sqrt\frac{\pi}{13}
\frac{q}{a^7}
,\quad
A_4=\frac{35q}{4a^5}
,\quad A_6=\frac{-21q}{2a^7}.
\ee

To further improve the efficiency of calculation, based on the Wigner-Eckart theorem \cite{Edmonds_2016},
Stevens proposed an equivalent operator approach \cite{Stevens_1952},
whose recipe is to replace
$\{x , y, z\}$ and $ ( x \pm i y )$ by $\{ L _ { x } ,\ L _ { y } ,\  L _ { z } \}$
and $L ^ { \pm }$, respectively.
Although $x$ and $y$ commute, $L_x$ and $L_y$ do not.
Thus, the symmetrical replacement is needed, for instance, $xy$ is replaced by the $(L_xL_y+L_yL_x)/2$.
Following the recipe,
the crystal field Hamiltonian Eq.~\eqref{eq:V46} can be converted to,
\be
H = B _ { 4 } ^ { 0 } \left[ O _ { 4 } ^ { 0 } + 5 O _ { 4 } ^ { 4 } \right]
+
B _ { 6 } ^ { 0 } \left[ O _ { 6 } ^ { 0 } - 21 O _ { 6 } ^ { 4 } \right],
\label{eq:cry_H}
\ee
where
\be
O_4^0=\left[35 L _ { z } ^ { 4 } - 30 L ( L + 1 ) L _ { z } ^ { 2 } + 3 L ^ { 2 } ( L + 1 ) ^ { 2 } + 25 L _ { z } ^ { 2 } - 6 L ( L + 1 )\right]\,,
\ee
\be
O_4^4=
\frac{1}{2}\left[\left( L ^ { + } \right) ^ { 4 } + \left( L ^ { - } \right) ^ { 4 }
\right]\,,
\ee
\be
\begin{split}
O_6^0&=
231L_z^6-(315L(L+1)
-735)L_z^4
+(105 (L(L+1))^2 - 525 L(L+1) + 294) L_z^2
\\& \quad
- 5 (L(L+1))^3 + 40 (L(L+1))^2 - 60 L(L+1)\,,
\end{split}
\ee
\be
\begin{split}
O_6^4&=
\frac{1}{4}
\left[ (11L_z^2-L(L+1)-38)(L_+^4+L_-^4)+(L_+^4+L_-^4)(11L_z^2-L(L+1)-38) \right]\,.
\end{split}
\ee
And the coefficients
\be
B_4^0=\frac{7}{16}\frac{-eq}{a^5}\cdot\beta\langle r^4\rangle,\quad
B_6^0=\frac{3}{64}\frac{-eq}{a^7}\cdot\gamma\langle r^6\rangle\,,
\ee
where
\be
\beta = \mp \frac { 3 ( 2 l + 1 - 4 S ) [ - 7 ( l - 2 S ) ( l - 2 S + 1 ) + 3 ( l - 1 ) ( l + 2 ) ] } { ( 2 l - 3 ) ( 2 l - 1 ) ( 2 l + 3 ) ( 2 l + 5 ) ( L - 1 ) ( 2 L - 1 ) ( 2 L - 3 ) }\,.
\ee
The minus (plus) sign is used for a shell less (more) than half full.
The values of $\gamma $ in different configurations are listed in TABLE 20 of \cite{AABB1970}.
Because $\beta<0$ and  $q<0$, $B^0_4<0$ for $3d^7$ electrons case.
And for the $d$-electrons, only $\mathcal{V}_4$ needs to be taken into account since $\mathcal{V}_6$ has much weaker effect on $d$-electrons.
Thus, the crystal field Hamiltonian for $d$-electrons follows
\be
H = B _ { 4 } ^ { 0 } \left[ O _ { 4 } ^ { 0 } + 5 O _ { 4 } ^ { 4 } \right]
\,.
\label{eq:cry_Hd}
\ee

\subsection{ Effective 1D spin-1/2 XXZ model }

For the octahedron $\rm CoO_6$, the crystal field Hamiltonian Eq.~\eqref{eq:cry_Hd} has already been diagonalized in the $\Gamma_{L=3}$ manifold Eq.~\eqref{eq:basis}.
As a result, the triplet $T_1$ has $360B_4^0$, the triplet $T_2$ has $-120B_4^0$ and the singlet $A_2$ has $-720B_4^0$.
{Because $B_4^0 < 0$ for $3d^7$ electrons,} the ground manifold of $\rm CoO_6$ is triplet $T_1$.
% \red{[Check $T_1$ triplet or triplet $T_1$.]}
If the energy gap between the ground state and the excited state is large,
we can use a pseudo angular momentum to describe the ground  manifold \cite{Fazekas1999,AABB1970}.
One can find that the matrix representation of ${L}$ in the $T_1$ manifold {can be effectively considered as it in} $p$-basis (i.e. $Y^1_1,\ Y_1^0,\ Y_1^{-1}$) with projection coefficient $-3/2$.
Thus, we can replace ${L}$ by $-3/2 \tilde L$ where $\tilde L=1$.
In this sense, $3d^7$ electrons in $T_1$ manifold behave as if they were $p$-electrons carrying $\tilde L=1$.

\begin{figure}[h]
\centering
\includegraphics[scale=0.4]{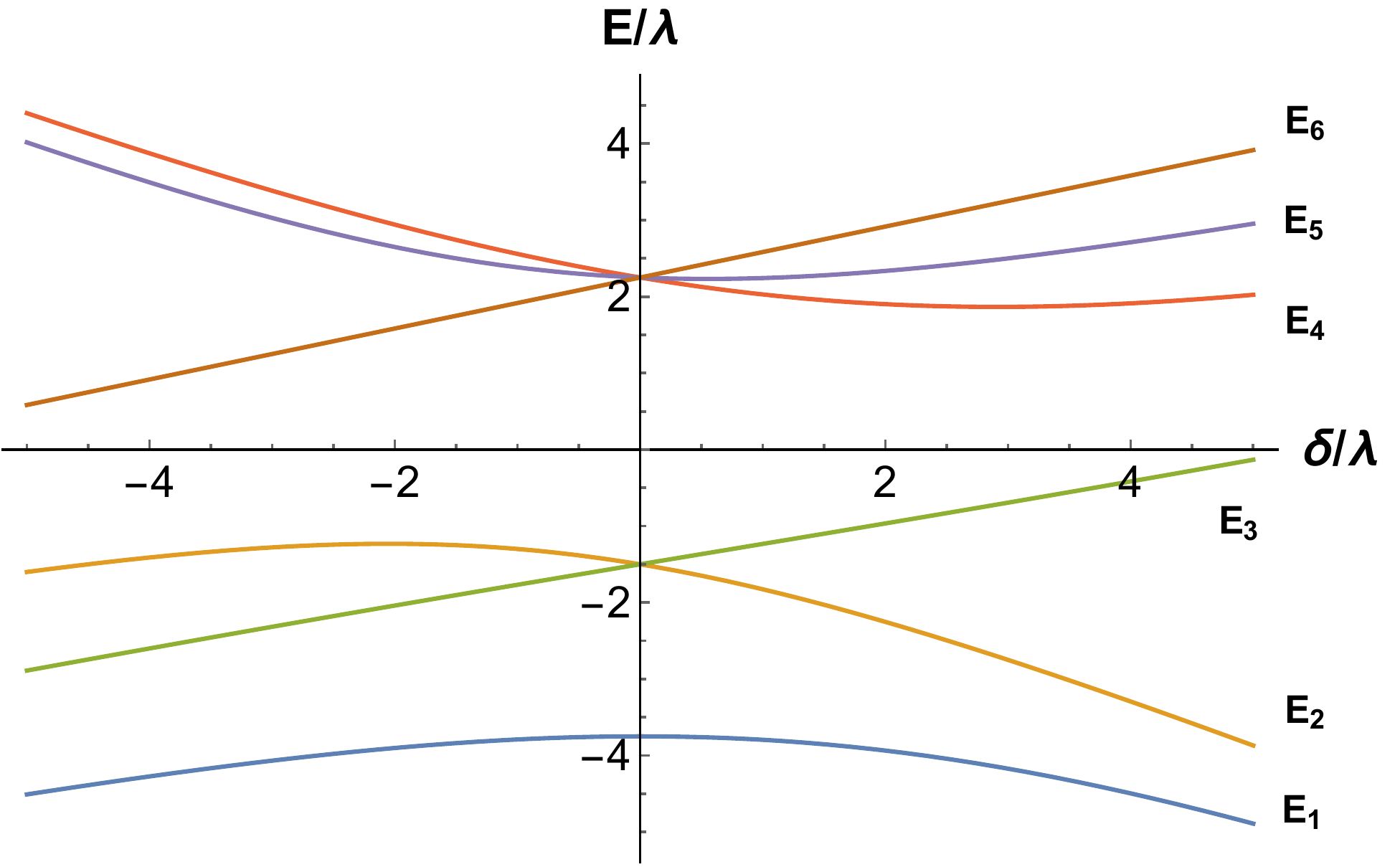}
\caption{
The six Kramers doublets due to crystal field splitting.
}
\label{fig:six_doublets}
\end{figure}

Recalling the total spin $S_{tot}=3/2$, the ground state is $3\times4=12$-fold degenerate,
which can be labeled as $|m_l, m_s\rangle$ where $m_l = -1, 0, 1 $ and $m_s = -3/2, -1/2, 1/2, 3/2 $
are components of $\tilde L$ and $S_{tot}$, respectively.
The SOC effect and the distortion play non-negligible roles in the materials we discuss,
which can lift the 12-fold degeneracy to certain degree.
The perturbation Hamiltonian follows
\be
H'=-\frac{3}{2}\lambda   {\bf\tilde L}\cdot\mathbf{S}_{tot}
-\delta(\tilde L_z^2-2/3)
\label{eq:Hper_SOC_dist}
\ee
where $\lambda$ and $\delta$ are coupling coefficient and distortion strength, respectively.
The Hamiltonian $H'$ has a diagonal form in states $|m_j\rangle$
where $m_j=m_l+m_s$ is the component of the total angular momentum $J=\tilde L+S_{tot}$.
Due to the time reversal symmetry of $H'$, $|\pm m_j\rangle$  are degenerate and form a Kramers doublet.
In total, there are $2\times5/2+1=6$ Kramers doublets whose energies as a function of $\delta/\lambda$ are shown in Fig.~\ref{fig:six_doublets}.
The ground state doublet with energy $E_1$ is given by:
\be
|\pm 1/2\rangle=
c_1|\mp1,\pm3/2\rangle
+c_2|0,\pm1/2\rangle
+c_3|\pm1,\mp1/2\rangle
\label{eq:gs_doublet}
\ee
where $c_1$, $c_2$ and $c_3$ are coeffecients dependent on $\delta/\lambda$.
When the ground state doublet is well separated from the second lowest one,
we can construct the matrix representation for the total spin $S_{tot}=3/2$
within the ground state doublet
\be
S^x_{tot}=\frac{\hbar}{2}
\begin{pmatrix}
0 & q\\
q & 0
\end{pmatrix}
,\quad
S^y_{tot}=\frac{\hbar}{2}
\begin{pmatrix}
0 & -iq\\
iq & 0
\end{pmatrix}
,\quad
S^z_{tot}=\frac{\hbar}{2}
\begin{pmatrix}
p & 0\\
0 & -p
\end{pmatrix}
\label{eq:S-1/2}
\ee
where $q=c_2^2+\sqrt{3}c_1c_3$ and $p=({3}c_1^2+c_2^2-c_3^2)/2$.
Then, in the lowest-lying Kramers doublet,
 the $S_{tot}=3/2$ can be effectively described by the pesudospin $S=1/2$,
\be
S_{tot}^x=qS^x, \quad
S_{tot}^y=qS^y, \quad
S_{tot}^z=pS^z.
\label{eq:Smapping}
\ee
In light of this, the pesudospin S=1/2 exhibits an anisotropy in $S^z$ component.
Similar results for trigonal distortion can also be obtained \cite{lines_magnetic_1963,shiba_exchange_2003,abragam_1951,achiwa_linear_1969}.

Based on superexchange mechanism
\cite{Anderson1959,Anderson1994,SAWATZKY1976},
the interaction of the neighbouring total spin is isotropic, thus the
effective spin-3/2 Hamiltonian is the XXX model,
\be
H_{XXX}
=J\sum_{i=1}^N \left(S^x_iS^x_{i+1}+S^y_iS^y_{i+1}
+S^z_iS^z_{i+1}
\right).
\label{eq:H_XXX}
\ee
However, for SCVO and BCVO, the low-energy physics can be effectively described by pseudospin $S=1/2$ instead of the total spin $S_{tot}=3/2$ [{\it c.f.} Eq.~\eqref{eq:Smapping}].
Thus, the effective spin-1/2 Hamiltonian for SCVO and BCVO follows
\be
H_{XXZ}=J\sum_{i=1}^N \left(S^x_iS^x_{i+1}+S^y_iS^y_{i+1}+\Delta S^z_iS^z_{i+1}
\right),
\label{eq:H_XXZ}
\ee
where $J > 0$ and
$\Delta={p^2}/{q^2}$ represents the Ising anisotropy.
In the BCVO and SCVO the energy gap between the ground state doublet and second lowest one is large \cite{bera_magnetic_2014,bera_string_dispersions_2020},
thus the effective Hamiltonian Eq.~\eqref{eq:H_XXZ} provides
a very good theoretical description for those materials,
which is detailedly discussed in Sec.~\ref{sec:excitations}.
With similar crystal field environment,
Co-based materials usually exhibits an Ising-anisotropy in the spin-spin interaction,
such as
the quasi-2D material $\rm Ba_3CoSb_2O_9$
\cite{Kamiya2018,Ghioldi2015}.
But this is not always the case \cite{Kim_2021}, since the origin of the spin-spin interaction is rather
complicated compared with the simple superexchange mechanism.

\subsection{ the Land\'{e} \texorpdfstring{$g$}{} tensor of \texorpdfstring{$\rm CoO_6$}{} screw chains }
\label{sec:Lande_g}

\begin{figure}[h]
\centering
\includegraphics[width=0.6\textwidth]{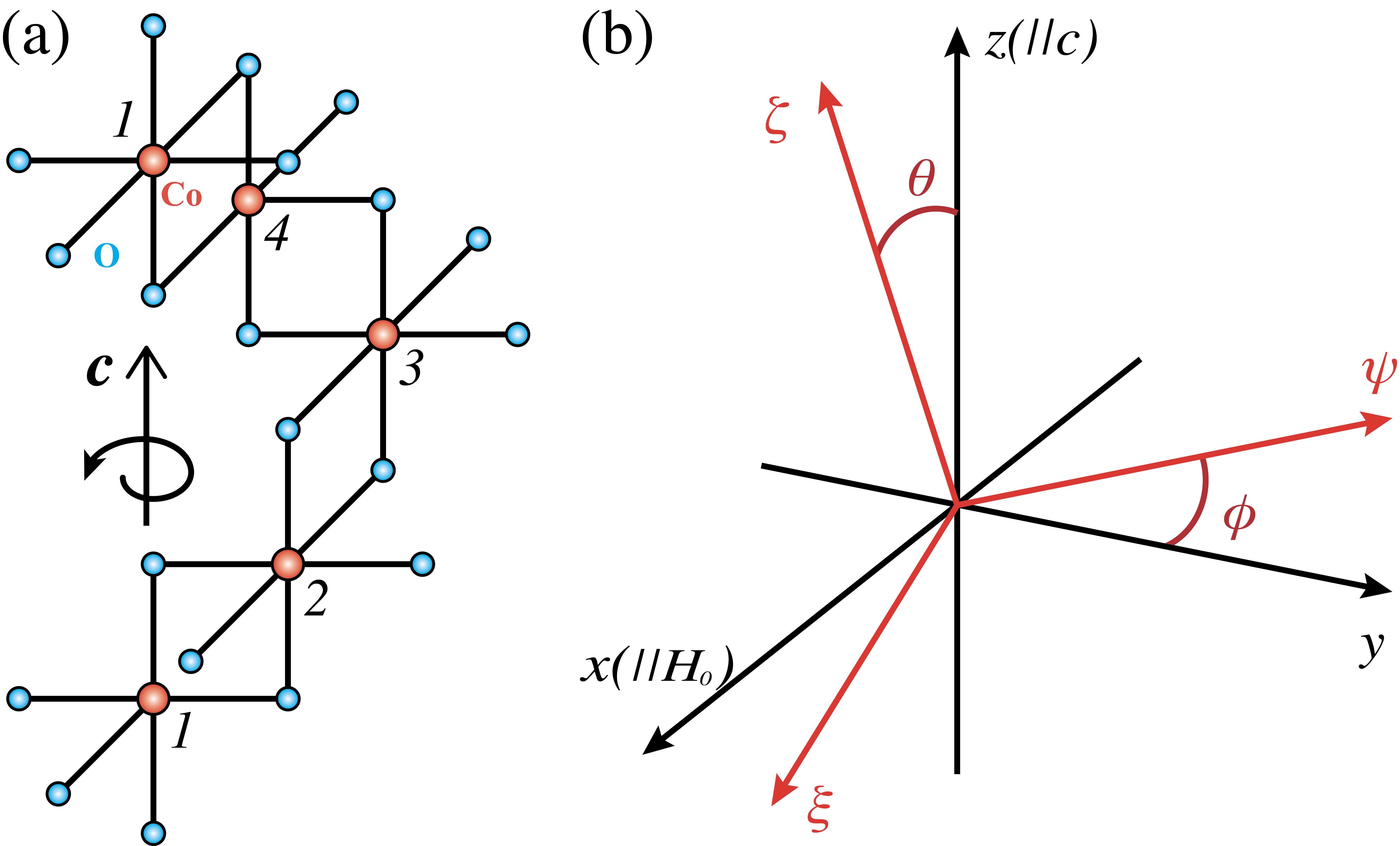}
\caption{Schematic view of
(a) $\rm CoO_6$ screw chain and
(b) the relation between the $xyz$- and $\xi \psi \zeta$-coordinate system.
}
\label{fig:ScrewCoOchain}
\end{figure}

In the following we shall analyze the Land\'{e} $g$ factor for the SCVO and BCVO materials \cite{bera_magnetic_2014,He_SCVO_2006,kimura_collapse_2013,faure_topological_2018}.
The SCVO and BCVO are quasi-1D material
with four-fold periodicity along the $\rm CoO_6$ screw chain [Fig.~\ref{fig:ScrewCoOchain} (a)].
Due to local slight inclinations, the local $\xi \psi \zeta$-coordinate system for a local
$\rm CoO_6$ octahedron has
a tilted angle from the right-handed $xyz$-coordinate system for the lab frame [Fig.~\ref{fig:ScrewCoOchain} (b)].
As such, the $\tilde g$ factor becomes a tensor from the view of $xyz$ frame where
in experiments the $z$-axis corresponds to the $c$-axis,
$x$-axis is set along the transverse field, and $y$-axis is perpendicular to the $xz$-plane.

The local $\xi \psi \zeta$-coordinate system is given by rotating the $xyz$-coordinate system around the $y$-axis by $\theta$ and then around the $z$-axis by $\phi$.
Thus, the effective $\tilde g$ factor expressed in the $xyz$-coordinate system follows
\be
\tilde g_{xyz}=
\begin{pmatrix}
{g_{xx}} & {g_{xy}} & {g_{xz}} \\
{g_{xy}} & {g_{yy}} & {g_{yz}} \\
{g_{xz}} & {g_{yz}} & {g_{zz}}
\end{pmatrix},
\label{eq:g_tensor}
\ee
where
\be
\begin{aligned}
g_{xx} &=\left(g_{\xi} \cos ^{2} \theta+g_{\zeta} \sin ^{2} \theta\right) \cos ^{2} \phi+g_{\psi} \sin ^{2} \phi, \\
g_{yy} &=\left(g_{\xi} \cos ^{2} \theta+g_{\zeta} \sin ^{2} \theta\right) \sin ^{2} \phi+g_{\psi} \cos ^{2} \phi, \\
g_{zz} &=g_{\xi} \sin ^{2} \theta+g_{\zeta} \cos ^{2} \theta, \\
g_{xy} &=\left(g_{\xi} \cos ^{2} \theta-g_{\psi}+g_{\zeta} \sin ^{2} \theta\right) \frac{\sin 2 \phi}{2}, \\
g_{yz} &=\left(g_{\zeta}-g_{\xi}\right) \sin \theta \cos \theta \sin \phi, \\
g_{xz} &=\left(g_{\zeta}-g_{\xi}\right) \sin \theta \cos \theta \cos \phi.
\end{aligned}
\ee
Here, $g_\xi$, $g_\psi$ and $g_\zeta$ are the values of $g$ factor in the $\xi\psi\zeta$-coordinate system.
In the screw chain, the angle $\phi$ shifts by $\pi/2$ if we change
$\rm Co^{2+}$ site to the next one along the chain, reflecting the four-fold periodicity of the screw structure.
We denote $\phi_1$ the angle between $y$-axis and $\psi$-axis for the site 1 in the screw chain.
Then the four-fold periodicity condition is $\phi=\phi_1+(j-1)\pi/2$ where $j=1,2,\ldots,N$ is the site index.
Thus, along the chain, the Zeeman effect of a transverse field $\bf H_0$ follows
\be
\begin{split}
{H}_{Zeeman}&=
\mu_B\sum_j \mathbf{S}_j\cdot \tilde{g}_{xyz}\cdot\mathbf{H}_0
\\&=
\mu_B\sum_j\left[ S_j^x g_{xx} \text{H}_0 +S_j^y g_{xy} \text{H}_0 +S_j^z g_{xz} \text{H}_0 \right],
\end{split}
\label{eq:zeeman_trans}
\ee
where
\be
\begin{aligned}
g_{xx}&=\left(g_{\xi} \cos ^{2} \theta+g_{\zeta} \sin ^{2} \theta\right) \cos ^{2} \left[ \phi_1+\frac{\pi}{2}(j-1) \right]+g_{\psi} \sin ^{2} \left[ \phi_1+\frac{\pi}{2}(j-1) \right],
\\
g_{xy}&=\frac{1}{2}\left(g_{\xi} \cos ^{2} \theta-g_{\psi}+g_{\zeta} \sin ^{2} \theta\right) \sin \left[2 \phi_1+\pi(j-1) \right],
\\
g_{x z}&=\frac{1 }{2}\left(g_{\zeta}-g_{\xi}\right) \sin 2\theta \cos \left[ \phi_1+\frac{\pi}{2}(j-1) \right].
\end{aligned}
\ee

\begin{figure}[h]
\centering
\includegraphics[width=0.8\textwidth]{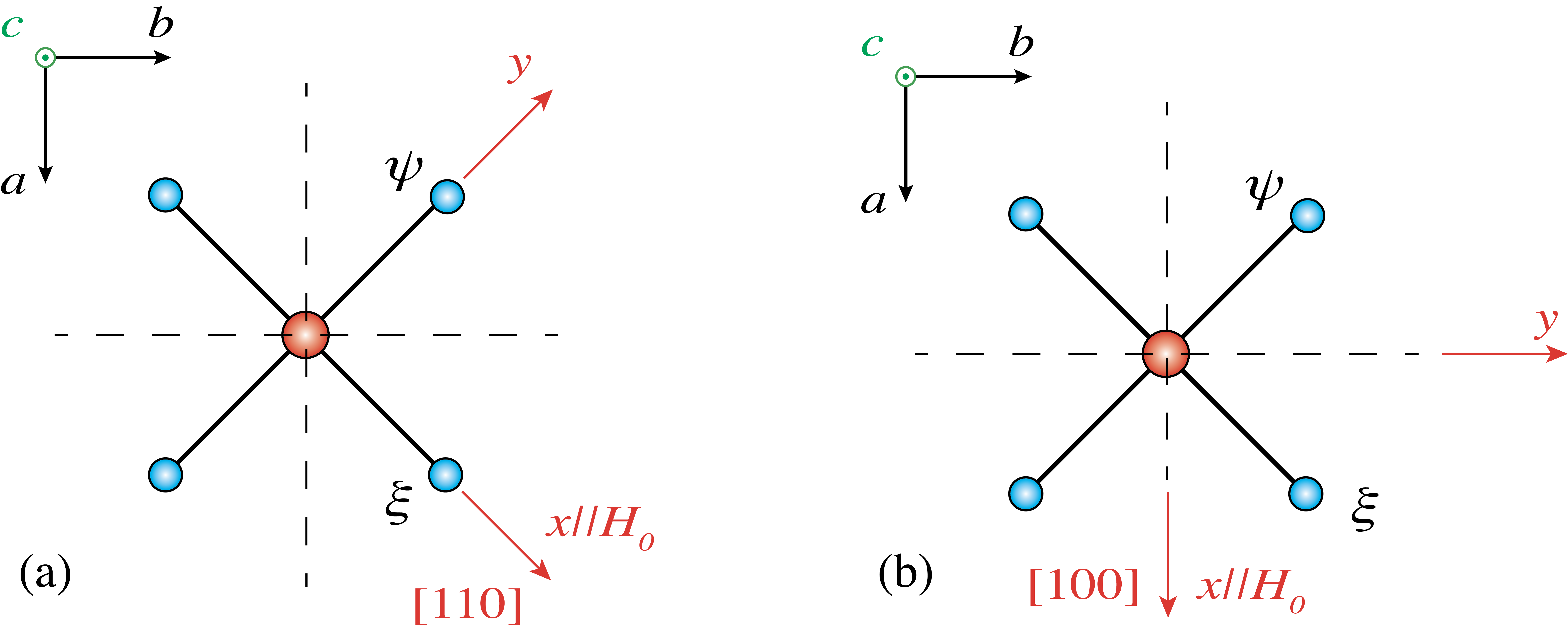}
\caption{Two cases with different directions of magnetic field.
}
\label{fig:2_H_direction}
\end{figure}

Next we present two cases with different directions of transverse field, i.e.  $\phi_1=0^\circ$ and $\phi_1=45^\circ$.
We should bear in mind that $\hat{x}//\mathbf{H}_0$.
In $\phi_1=0^\circ$ case, namely, $\mathbf{H}_0$ is along $[110]$ direction [Fig.~\ref{fig:2_H_direction} (a)], then
the components of $\tilde g$ factor become
\be
\begin{aligned}
g_{xx}&=\left(g_{\xi} \cos ^{2} \theta+g_{\zeta} \sin ^{2} \theta\right) \cos ^{2} \left[ 0+\frac{\pi}{2}(j-1) \right]+g_{\psi} \sin ^{2} \left[ 0+\frac{\pi}{2}(j-1) \right],
\\
g_{xy}&=\frac{1}{2}\left(g_{\xi} \cos ^{2} \theta-g_{\psi}+g_{\zeta} \sin ^{2} \theta\right) \sin \left[0+\pi(j-1) \right]=0,
\\
g_{xz}&=\frac{1 }{2}\left(g_{\zeta}-g_{\xi}\right) \sin 2\theta \cos \left[ 0+\frac{\pi}{2}(j-1) \right],
\end{aligned}
\label{eq:g_110}
\ee
which indicates that
the transverse field along [110] direction can induce a four-periodic field but without staggered field.

In $\phi_1=45^\circ$ case, namely, $\mathbf{H}_0$ is along $[100]$ direction [Fig.~\ref{fig:2_H_direction} (b)], then
the components of $\tilde g$ factor become
\be
\begin{aligned}
g_{xx}&=
\frac{1}{2}
\left[g_{\xi}\cos^2\theta+g_{\zeta}\sin^2\theta+g_{\psi}\right],
\\
g_{xy}&=
(-1)^{j} \frac{1}{2}\left(g_{\xi} \cos ^{2} \theta-g_{\psi}+g_{\zeta} \sin ^{2} \theta\right),
\\
g_{xz}&=\frac{1}{2}\left(g_{\zeta}-g_{\xi}\right) \sin 2\theta \cos \left[ \frac{\pi}{4}(2j-1) \right],
\end{aligned}
\label{eq:g_100}
\ee
which indicates that
the transverse field along [100] direction can induce staggered and four-periodic fields.
One can obtain a similar result for the [010] direction.
% If the field ${\bf H}_0$ is along [010] direction, there is only a $\pi/2$ shift for $\phi_1$ and the results is similar to [100] case.
Now we are ready to discuss magnetic excitations
in the 1D spin-1/2 XXZ model with various external fields,
as well as their
experimental realization in the BCVO and SCVO materials.

\section{Magnetic Excitations in the 1D Spin-1/2 XXZ Model with Various External Fields} \label{sec:excitations}

Depending on the directions of the applied external field, the magnetic excitations
in the spin-1/2 XXZ chain exhibit rich emergent phenomena and exotic physics,
e.g., Tomonaga-Luttinger liquid \cite{tomonaga_remarks_1950,Luttinger_1960,Luther_TLL_1975},
spinon \cite{FADDEEV_spinwave_1981,faddeev_spectrum_1984,Muller_spinon_1981,Karbach_spinon_1997,Bougourzi_spinon_1996,Bougourzi_spinon_1998,caux2008},
string \cite{bethe1931,takahashi_1D_1971,Gaudin_XXZ_1971,Taka_suzuki_XXZ_1972,kohno_string_dynamically_2009},
and $E_8$ particles \cite{a_b_zamolodchikov_integrals_1989,jianda_E8_2014,DELFINO_1995}.
To help measure or probe those exotic magnetic excitations,
a study on the spin dynamical structure factor (SDSF) becomes a must,
which can reveal weight distributions and dispersions of
various magnetic excitations on one hand, and give a direct guidance for their experimental
realization on the other hand \cite{Negele1988,chaikin_lubensky_1995,Zhu_MTCMM_2005}.
At zero temperature, the SDSF follows the Fourier transform of the
spin correlation function in space and time,
\be
S^{a\bar{a}}(q,\omega)
=
\frac{1}{N}\sum_{jj'}^N e^{- iq(j-j')}
\int^{+\infty}_{-\infty}dt e^{i\omega t}
\langle GS | S_j^a(t)S_{j'}^{\bar{a}}(0)| GS \rangle
\ee
where $a\in \{-,+,z\}$ and $|GS\rangle$ is the ground state.
Using the Lehmann representation,
it reduces to a single summation over a complete basis set,
\be
S^{a\bar{a}}(q,\omega)=
2\pi\sum_{\alpha}
|\langle GS |S_q^a|\alpha \rangle|^2\delta(\omega-\omega_\alpha)
\label{eq:SDSF}
\ee
where $\omega_\alpha=E_\alpha-E_0$ is the energy difference between
excited state $|\alpha\rangle$ and ground state $|GS\rangle$.
In Eq.~\eqref{eq:SDSF},
the spectral weight
$|\langle GS |S_q^a|\alpha \rangle|^2$
describes the transition probability between $|GS\rangle$ and $|\alpha\rangle$ after an external perturbation (the incoming photon or neutron) which couples linearly to the $S^a_q$ operator \cite{Negele1988,chaikin_lubensky_1995,Zhu_MTCMM_2005}.
Therefore, based on
spectral weight distribution $|\langle GS |S_q^a|\alpha \rangle|^2$
from different excitations $|\alpha\rangle$
in the energy-momentum space,
we can carry out detailed comparison
between theoretical calculation and experimental measurement
for recognizing possible realizations of various types
of magnetic excitations.
In the following, we will discuss the magnetic excitations in the spin-1/2 XXZ chain with
various external fields and their experimental realizations.

\subsection{ Zero field }

\begin{figure}[h]
\centering
\includegraphics[width=0.7\textwidth]{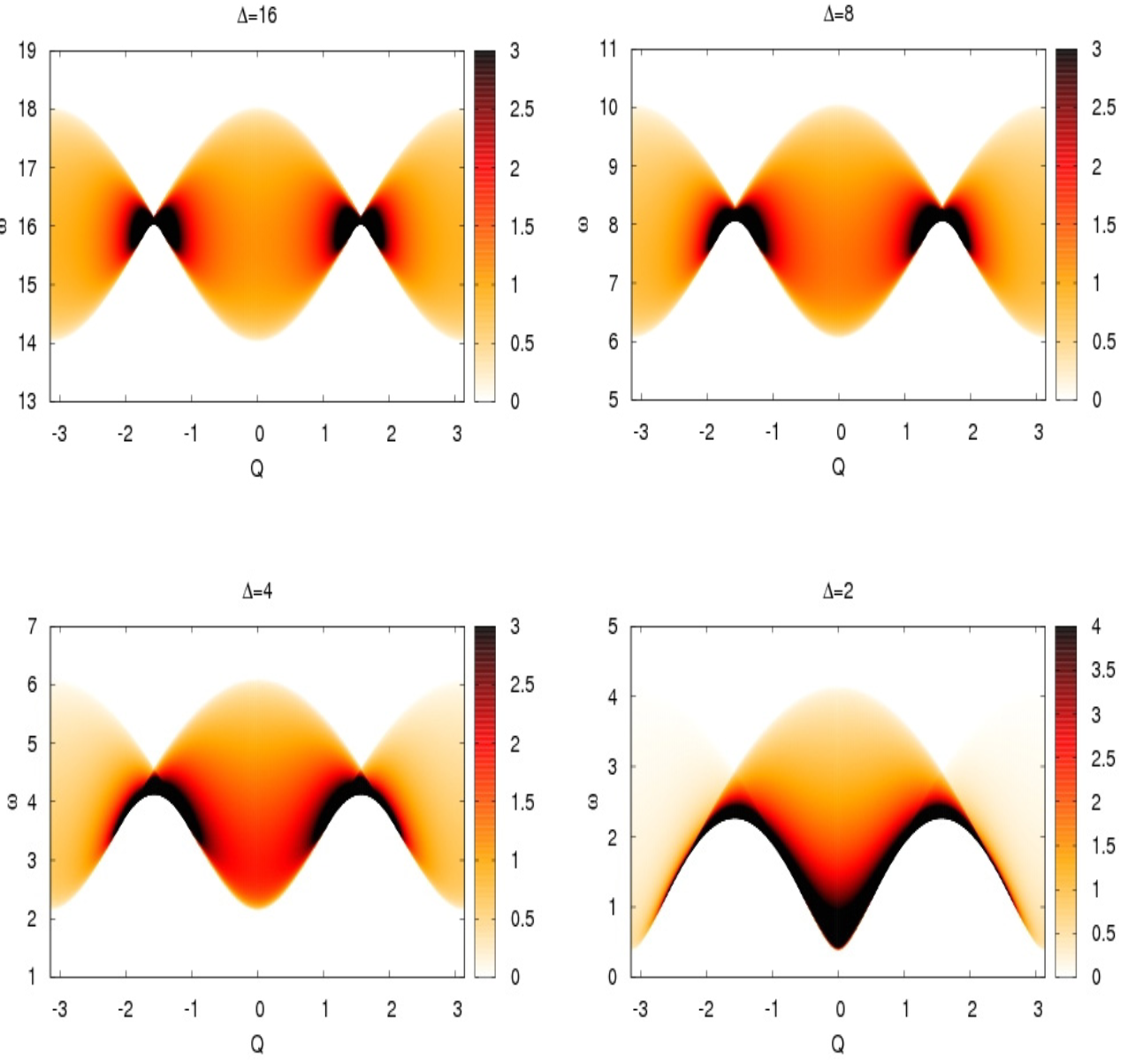}
\caption{
The density plots for $S^{-+}$ at different anisotropy $\Delta$. From \cite{caux2008}. }
\label{fig:spinon_mp}
\end{figure}

\begin{figure}[h]
\centering
\includegraphics[width=0.55\textwidth]{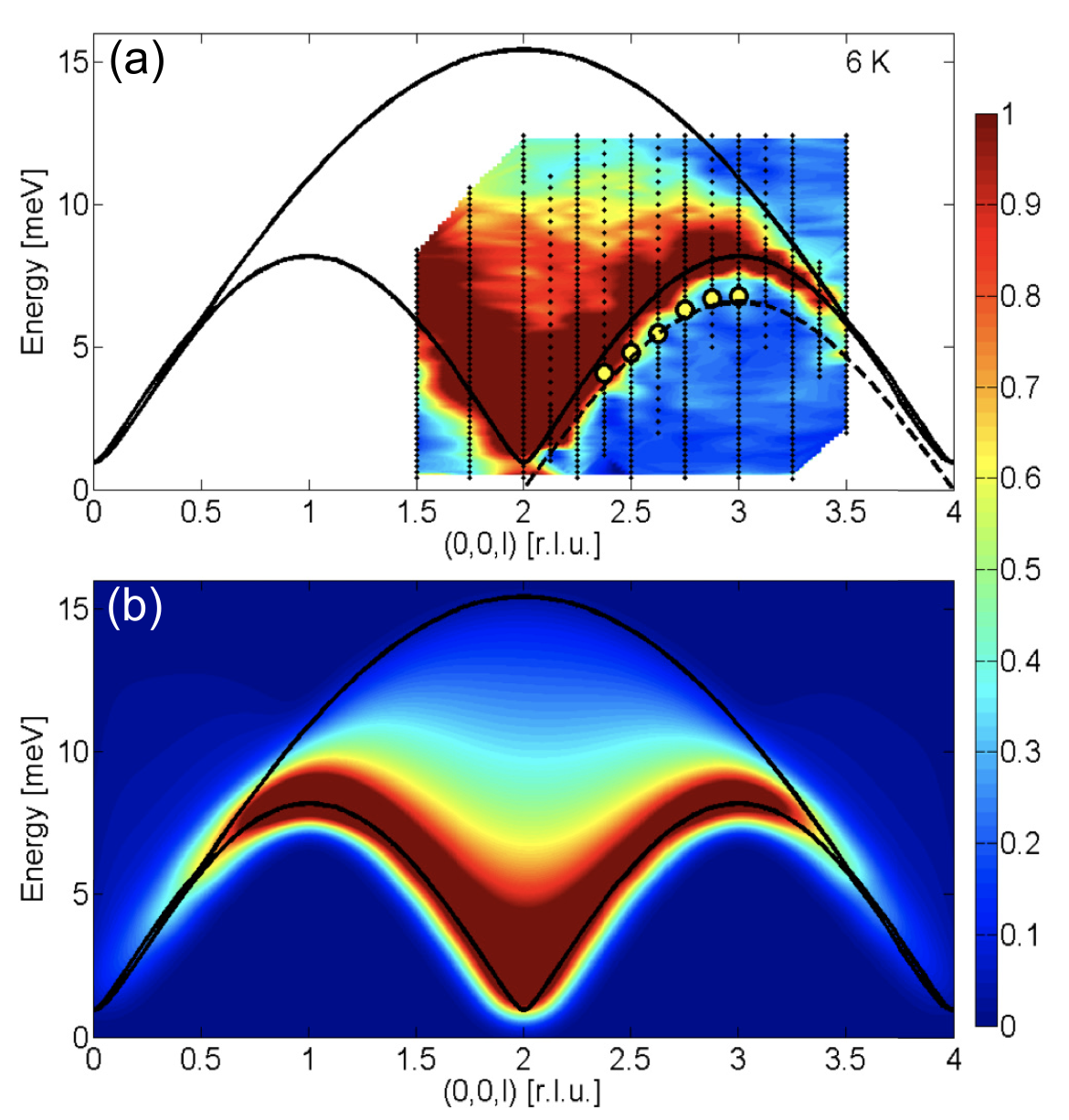}
\caption{Dynamical spectrum of spinons. (a) INS measurement in SCVO and (b) Bethe ansatz calculation.
The theoretical dispersion of the Villian mode is represented by the dashed line
and the yellow points are used to highlight experimental peaks in (a).
From \cite{bera_spinon_2017}.}
\label{fig:spinon_INS_BA}
\end{figure}

At zero field, ground state of the 1D spin-1/2 XXZ model [Eq.~\eqref{eq:H_XXZ}]
accommodates a long-range AFM ordering
with an energy gap dependent on the Ising anisotropic parameter $\Delta$.
Above the gap, spin-flip excitations fractionalize
into pairs of spinons with fractional quantum number $S=1/2$ \cite{FADDEEV_spinwave_1981}.
By studying the SDSF of the system, we are able to
explore how the spinons influence the dynamics of the system.
For the $S^{-+}(q,\omega)$ [{\it c.f.} Eq.~(\ref{eq:SDSF})], the two-spinon
excitations provide a gapped continuum spectrum \cite{caux2008}, as shown in Fig.~\ref{fig:spinon_mp}.
The two-spinon excitations are
a small fraction of the total number of states,
however, they carry almost {100\%} spectral weight in $S^{-+}$ channel if $\Delta\ge2$.
When $\Delta < 2$ multi-spinon states like four-spinons excitations
play a more and more important role with decreasing $\Delta$ toward
the isotropic limit ($\Delta =1$) \cite{caux2008,Caux_2006}.
In $S^{zz}(q,\omega)$ channel, two-spinon excitations have similar results as
in $S^{-+}(q,\omega)$ \cite{castillo_exact_2020}.
The analytical SDSF provides a concrete ground
for the experimental realization.
Indeed, INS measurements
on the quasi-1D material SCVO directly
observe the expected excitation spectrum.
In SCVO, the long-range AFM order appears below
its N\'eel temperature $T_N=5.2$ K.
1D physics is expected to take over dominancy when $T > T_N $,
where SCVO can be effectively
described by the 1D spin-1/2 XXZ model Eq.~\eqref{eq:H_XXZ}.
The INS measurements at 6 K reveal a gapped scattering spectrum [Fig.~\ref{fig:spinon_INS_BA}] with
excellent agreement between the theoretical and experimental results \cite{bera_spinon_2017}.
Additionally,
Fig.~\ref{fig:spinon_INS_BA}(a)
also exhibits resonance peaks
caused by thermal fluctuation
and referred to as the Villian modes \cite{Villain_1975}.
These modes are firstly observed in $\rm CsCoBr_3$ by INS experiments
and can be explained by the scattering between two domain-wall states
\cite{Nagler_villian_PRL_1982,Nagler_villian_PRB_1983,James_XXZ_FiniteT_2009}.
Next, we shall discuss the magnetic excitations in 1D spin-1/2 XXZ model
with the presence of various external magnetic fields.

\subsection{Longitudinal field}

With a longitudinal field $h$, the Hamiltonian becomes
\be
H=
H_{XXZ}
-h\sum_{i=1}^N S^z_i,
\label{eq:H_XXZ_lh_sec4}
\ee
where the time reversal symmetry is broken but the $U(1)$ symmetry is preserved.
If the longitudinal field is small such that the gap holds, i.e.,
the AFM ground state maintains,
there is no magnetization.
The magnetization starts to develop when $h$ is tuned above the field threshold $h_c(\Delta)$ where
the gap closes, and then the system enters the quantum critical (gapless) regime.
In this region,
the spinon can no longer serve as a good quasi-particle to describe the excitations
in the model.
Instead, the dynamics is dominated by
fractional excitations known as
``psinon-psinon'' (PP) and ``psinon-antipsinon'' (PAP)
\cite{Karbach_pp_pap_2002,karbach2000III},
as well as exotic string excitations \cite{yang_string_one-dimensional_2019}.
For observing those exotic states, an SDSF study needs to be carried out
in order to probe the excitations and quantify their spectra contributions.

\begin{figure}[h]
\centering\includegraphics[scale=0.6]{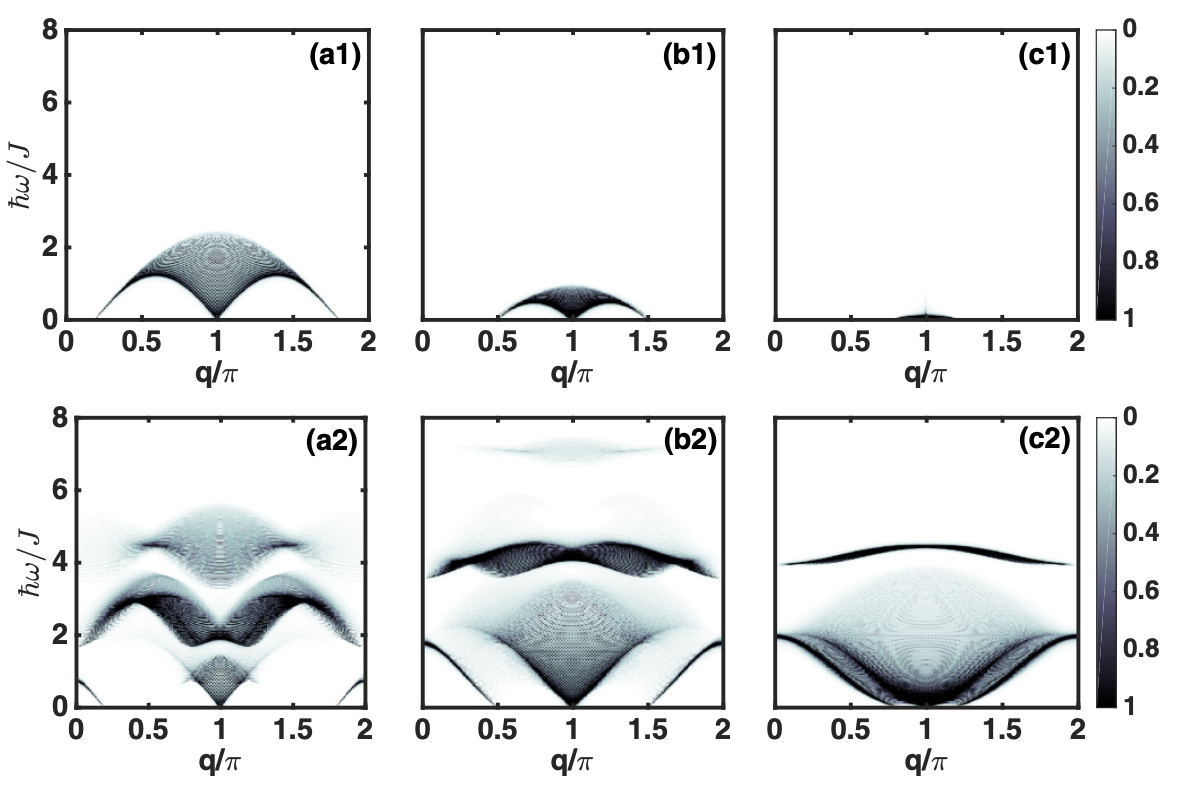}
\caption{
The density plots for $S^{-+}$ (top row) and $S^{+-}$ (bottom row) of
the 1D spin-1/2 XXZ model,
where the magnetization is 20\% (a1, a2), 50\% (b1, b2) and 80\% (c1, c2) of full polarization.
From \cite{yang_string_one-dimensional_2019}.}
\label{fig:DSF_XXZ}
\end{figure}

In the $S^{-+}(q,\omega)$ channel, fractional excitations PP have a similar SDSF shape
as that of spinon [Fig.~\ref{fig:DSF_XXZ}] at small magnetization and disappears
when the magnetization becomes saturated \cite{yang_string_one-dimensional_2019}.
%In contrast, contributions from PAP
%diminishes in the zero field limit and recover the spin-wave excitations in the
%limit of full polarization.
In both zero and nonzero field cases, these fractional excitations dominate almost the whole spectrum
implying negligibly small contribution from the
string excitations in the $S^{-+}(q,\omega)$ channel.
In a sharp contrast, for the $S^{+-}(q,\omega)$ channel [Fig.~\ref{fig:DSF_XXZ}],
contributions from low-energy factional excitations PAP
diminish in the zero field limit and recover the spin-wave excitations in the
limit of full polarization.
String excitations take over the dominance at the low-magnetization regions,
and tend to vanish when the system approaches full polarization.
At relatively low magnetization,
three well-separated continuums indicate three
different contributions from fractional PAP,
2-string and 3-string excitations.
The theoretical progress provides a concrete guidance to directly probe the
string excitations in real materials.

\begin{figure}[h]
\centering
\includegraphics[width=0.75\textwidth]{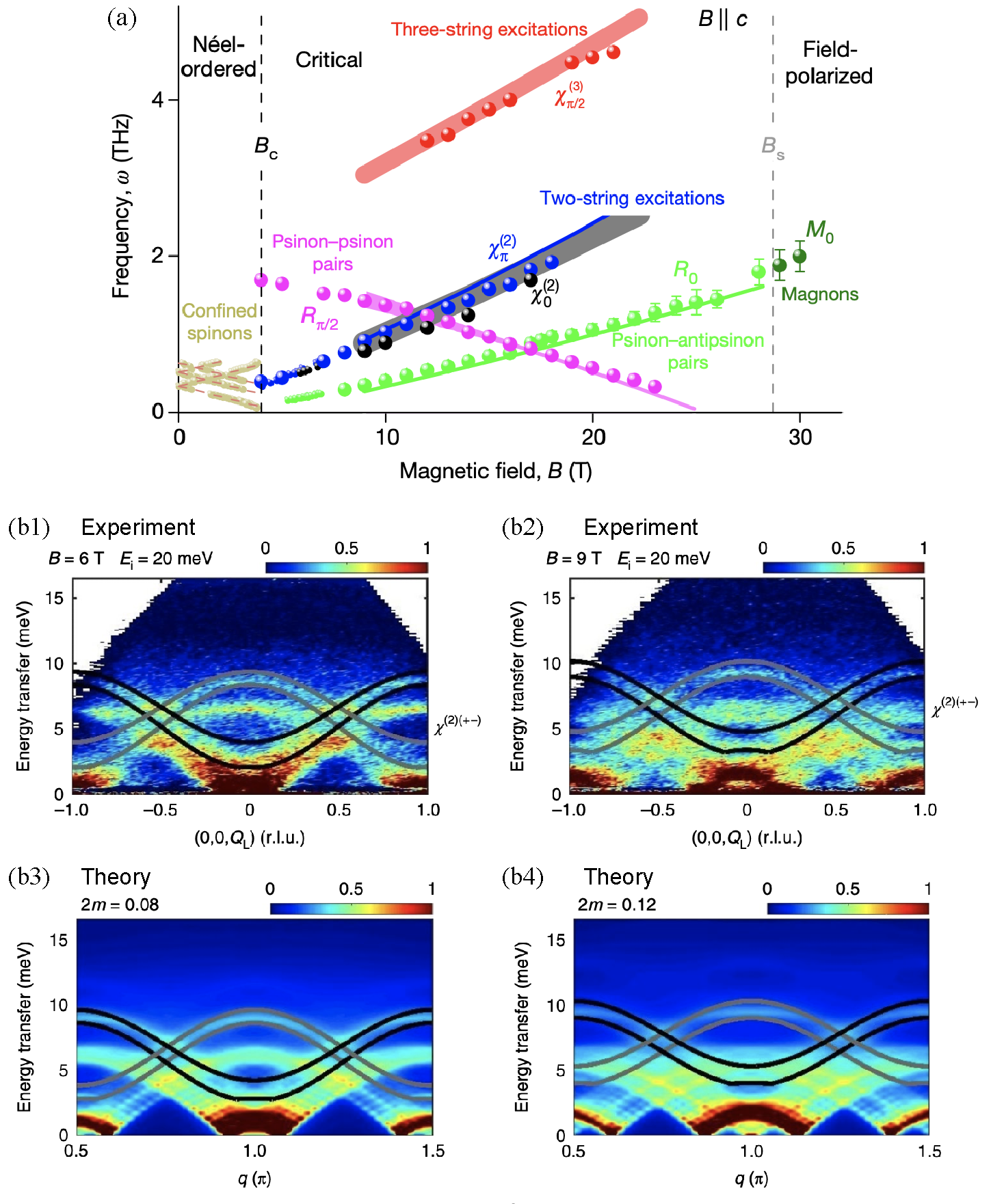}
\caption{(a) THz spectroscopy; (b1, b2) INS measurement of SCVO;
and (b3, b4) Bethe ansatz calculations.
From \cite{wang_string_experimental_2018,bera_string_dispersions_2020}.
}
\label{fig:THz_INS}
\end{figure}

In spite of the intriguing many-body nature of the excitations, especially the string excitations,
it is a great challenge to directly detect or
realize these exotic excitations in real materials
\cite{Imambekov_1D_2012,kohno_string_dynamically_2009,Pereira_edge_2008,Pereira_spectral_2009,
Caux_comput_2005,Caux_computation_2005,Shashi_Nonuniversal_2011,Ganahl_Observation_2012}.
In 2017, a silver lining appears, which suggests that the field-induced quantum critical region
of the 1D spin-1/2 XXZ model Eq.~\eqref{eq:H_XXZ_lh_sec4} is
a promising region to directly observe the string excitations \cite{yang_string_one-dimensional_2019}.
Following the suggestion, a high-resolution THz spectroscopy
measurement observes the string excitations in SCVO at the zone cneter
for the first time \cite{wang_string_experimental_2018}.
Under the concrete theoretical guidance, 2- and 3-string excitations
as well as low-energy fractional excitations (PP and PAP) are probed [Fig.~\ref{fig:THz_INS} (a)].
In 2020, the dispersion relation of the string excitations
over the full Brillouin zone is obtained via the INS measurement on the same material \cite{bera_string_dispersions_2020}.
Fig.~\ref{fig:THz_INS} (b) shows that the dispersion for the 2- and 3-string excitations
appears in the intermediate (2 - 15 meV at 6 T) and high (4 - 15.5 meV at 9 T)
energy regions, respectively.
In both experiments, the obtained excitation spectra and their magnetic field dependencies
are perfectly consistent with theoretical predictions,
closing the long-time open problem to directly observe the string excitations.

\subsection{Transverse field}

With presence of a transverse field the XXZ Hamiltonian
becomes
\be
H=
H_{XXZ}+H_x\sum_{i=1}^N S^x_i\,.
\label{eq:H_XXZ_Hx}
\ee
With increasing field, a quantum phase transition emerges, which
falls in the class of TFIC universality \cite{dmitriev_1D_2002}.
For realizing the TFIC universality in real material,
distinctive criticality is desired for discerning the TFIC universality from others.
It is found that the Gr\"uneisen ratio, directly related to the magnetocaloric effect,
exhibits a very unique quantum critical scaling. The Gr\"uneisen ratio either
approaches to a constant or divergence when the QCP
of the TFIC universality is accessed via decreasing temperature or tuning field,
respectively ~\cite{Jianda_crossover_2018}.
The critical behaviors are different
from general quantum criticality where the Gr\"uneisen ratio in general approaches to
divergence regardless how to access the QCP \cite{Zhu_Universally_2003}.
As such measurements of Gr\"uneisen ratio near a QCP of the real material
can serve as a smoking gun to justify whether the universality falls
in the TFIC universality.

As discussed in Sec.~\ref{sec:Lande_g}, in the BCVO and SCVO,
when the applied field is along [100] direction,
there are two field-induced terms [{\it c.f.} Eq.~\eqref{eq:zeeman_trans}] in the Hamiltonian,
\be
\begin{split}
H&=
H_{XXZ}
- H_x \sum_{i=1}^N\left\{ S_{i}^{x}
+h_y(-1)^{i}S_{i}^{y}
+h_zS^{z}_{i}
\cos\left[{\pi(2i-1)}/{4}
\right]
\right\}
\,,
\end{split}
\label{eq:H_XXZ_Hxzy}
\ee
where $h_y \approx 0.4$ and $h_z \approx 0.14$ are the
internal-induced staggered and four-periodic fields (reduced by $H_x$), respectively.
And if the applied field is along [110] direction, only the four-periodic field with
slightly different form is present [{\it c.f.} Eq.~\eqref{eq:g_110}].
Although the Hamiltonian Eq.~\eqref{eq:H_XXZ_Hxzy} becomes more involved compared with Eq.~\eqref{eq:H_XXZ_Hx},
it still preserves the TFIC universality when the system is tuned to its QCP.
It is found that
the four-periodic perturbation term only slightly changes the location of
the QCP,
while
the staggered-field can significantly reduce
the field strength to access the QCP \cite{Zou_universality_2019,faure_topological_2018}.
% and preserves the TFIC universality.
Thus, from above theoretical analysis,
both BCVO and SCVO effectively described by 1D spin-1/2 XXZ model,
can accommodate the TFIC universality
via tuning the transverse field properly.

\begin{figure}[h]
\centering
\includegraphics[width=0.9\textwidth]{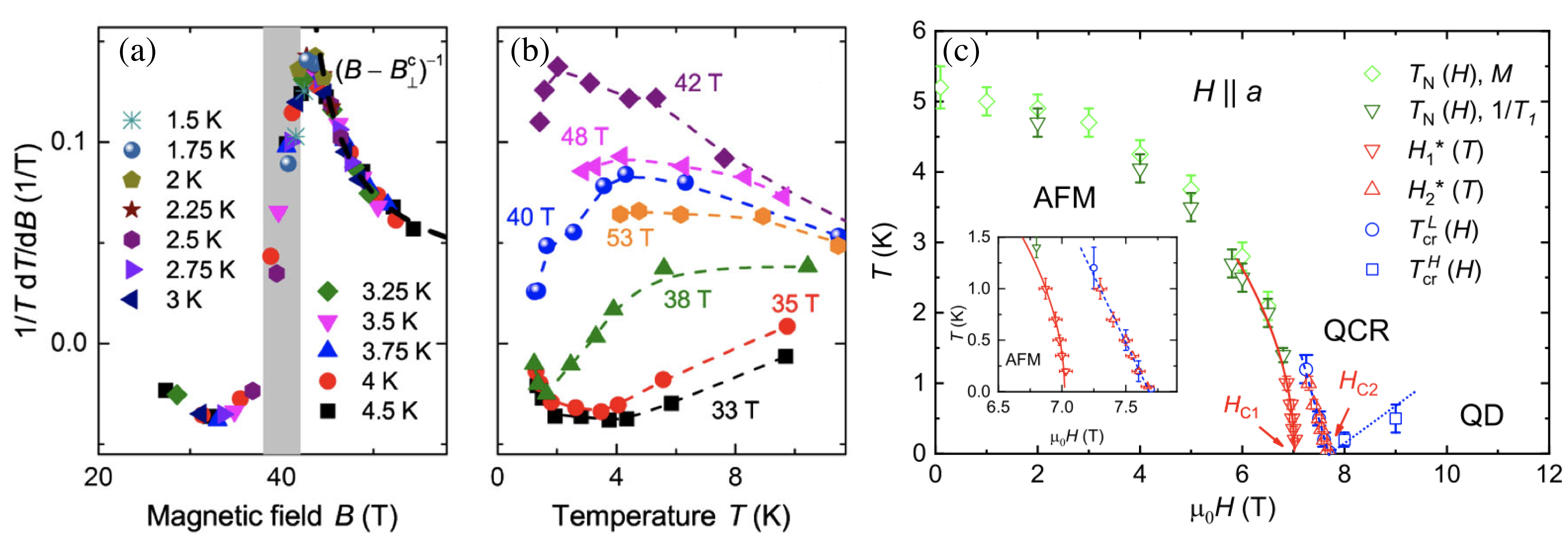}
\caption{
The Gr\"uneisen parameter for BCVO vs. transverse field (a) and temperature (b).
Phase diagram of SCVO vs. transverse field (c).
From \cite{cui_tfic_quantum_2019, wang_tfic_quantum_2018}.
}
\label{fig:TFIC}
\end{figure}

As expected, by applying a magnetic field along [110] direction,
the TFIC universality is realized in BCVO where the critical field is around 40 T \cite{wang_tfic_quantum_2018}.
The measured Gr\"uneisen ratio is divergent when the transverse field approaches 40 T
but converges when the temperature decreases at 40 T [Fig.~\ref{fig:TFIC} (a) and (b)],
consistent with the theoretical analysis~\cite{Jianda_crossover_2018,wang_tfic_quantum_2018}.
If the magnetic field is along [100] direction, much lower critical fields can be obtained in both SCVO and BCVO
\cite{Zou_universality_2019,cui_tfic_quantum_2019,zou_e_8_2021}.
For instance, in SCVO, two QCPs at $H_{C1} = 7$ T and $H_{C2} = 7.7$ T are determined by NMR experiment \cite{cui_tfic_quantum_2019}. The former one is a (3+1)D Ising critical point
while the later one falls in the class of TFIC universality [Fig.~\ref{fig:TFIC} (c)].
Near the critical field $H_{C2}$, the critical exponent of transverse field is found to be consistent with the 1D TFIC universality class \cite{sachdev_2011,cui_tfic_quantum_2019}.

\subsection{More field and the \texorpdfstring{$E_8$}{} physics}

\begin{figure}[h]
    \centering
	\includegraphics[width=0.72\textwidth]{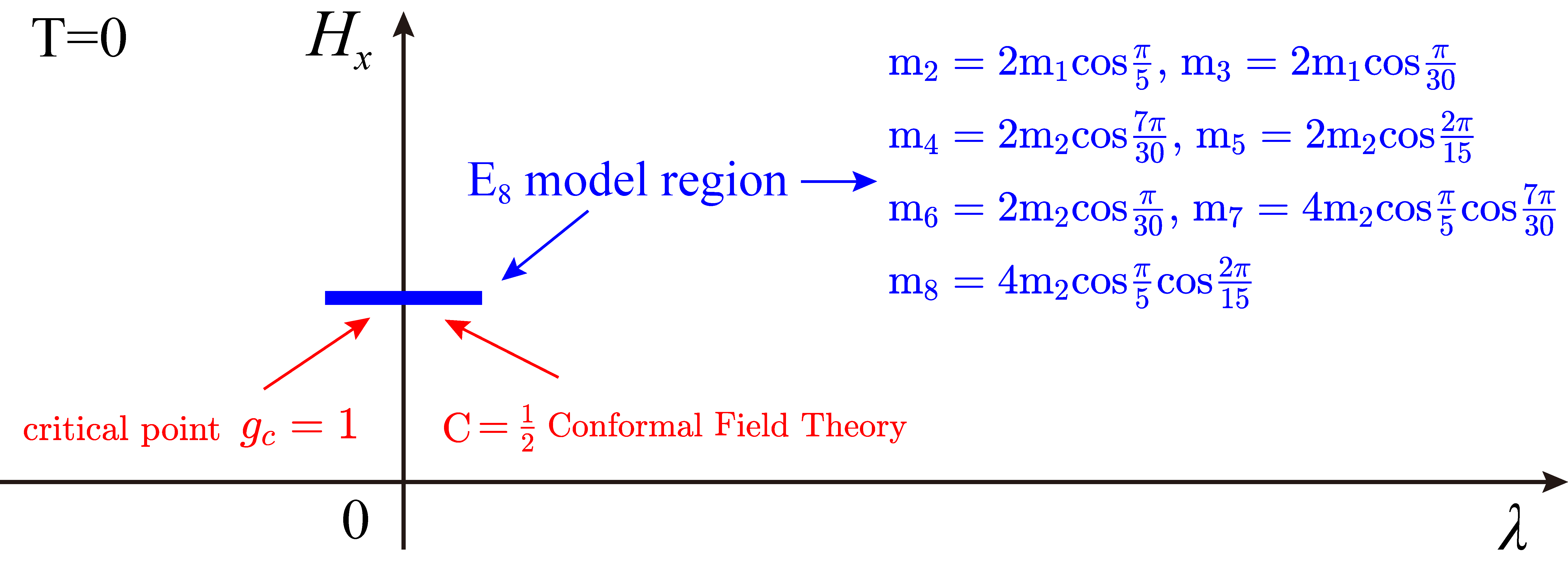}
	\caption{
	The quantum $E_{8}$ integrable model emerges in the region of blue solid line.
    Masses of the eight particles are expressed in units of the lightest two particles' masses $m_{1}$ and $m_{2}$.
    From~\cite{xiao_cascade_2021}.
	}
	\label{fig:E8_phasedia}
\end{figure}

In the vicinity of QCPs, the low energy excitations in the infrared (IR) limit could be described by conformal field theory (CFT) due to the scaling invariant and large conformal symmetry. Moreover, with perturbation deformation from relevant primary field of corresponding CFT, the original physics in the IR limit will be greatly influenced. For instance, it may turn to massive field theory associated with certain Lie algebra, dubbed as affine toda field theory (ATFT)~\cite{BRADEN1990689,MUSSARDO1992215}.
In the scaling limit, the TFIC universality
can be described by a CFT with central charge $c=1/2$.
With an additional small longitudinal-field perturbation,
an integrable massive quantum field theory (QFT)
dubbed as quantum $E_{8}$ model emerges.

The quantum $E_8$ integrable model contains eight different types of massive particles
with their scattering fully described by the $E_8$ exceptional Lie algebra~\cite{a_b_zamolodchikov_integrals_1989,jianda_E8_2014,DELFINO_1995,xiao_cascade_2021}.
The $E_8$ Hamiltonian follows
\be
H_{E_{8}}=H_{c=1/2}+\lambda\int\sigma(x)dx,
\label{eq:E8}
\ee
with the $c=1/2$ CFT Hamiltonian $H_{c=1/2}$ and
the intensity of the small longitudinal field in the scaling limit $\lambda$.
The mass of the lightest quasiparticle $m_{1}=C\lambda^{8/15}$, $C\approx 4.40490858$~\cite{DELFINO_1995,fateev}.
And $\sigma(x)$ is the spin density operator,
a relevant field in the perturbed CFT~\cite{a_b_zamolodchikov_integrals_1989,DELFINO_1995}.
Fig.~\ref{fig:E8_phasedia} shows the phase diagram and massive excitations of the $E_{8}$ model.

\begin{figure}[h]
    \centering
	\includegraphics[width=0.5\textwidth]{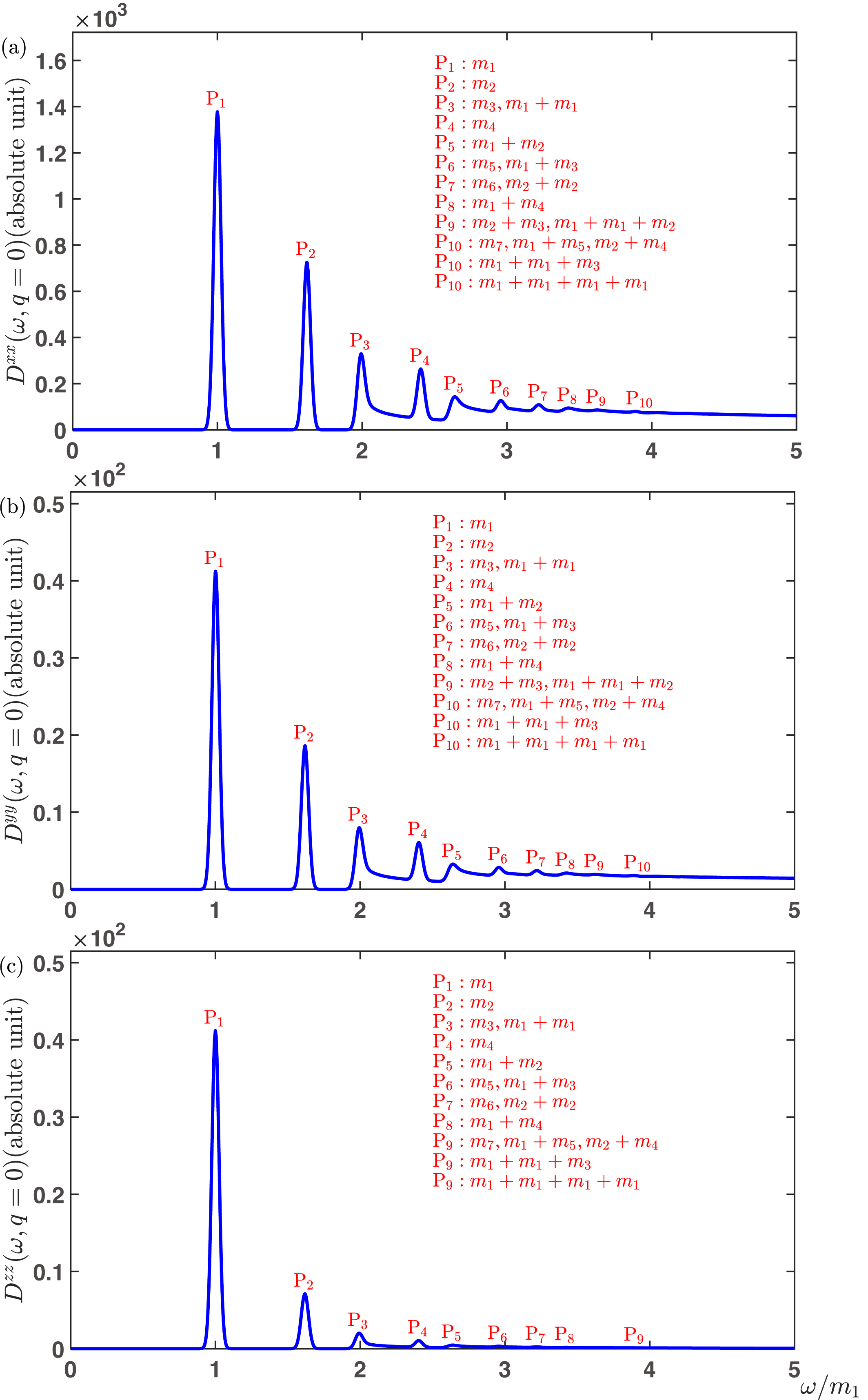}
	\caption{Analytical calculation of DSF with boradening of $0.05m_{1}$ has been exhibited. (a), (b) and (c) are corresponding to $S^{xx}(\omega)$, $S^{yy}(\omega)$ and $S^{zz}(\omega)$ respectively. All the contributions of different peaks have been highlighted inside the figures, from ~\cite{xiao_cascade_2021}.}
	\label{fig:E8_DSF_analytic}
\end{figure}

The eigenstates of ATFTs satisfy Faddeev-Zamolodchikov algebra and can be described by exact S-matrix theory~\cite{ZAMOLODCHIKOV1979253,BRADEN1990689}. For solving the spectrum of ATFTs, a programmable method dubbed as conformal bootstrap is proposed by Cardy and Mussardo~\cite{CARDY1990387}, which could generate expressions of form factor for certain operator.
Later, the form factor theory of the $E_{8}$ model is further developed by Zamolodchikov, Mussardo and Delfino~\cite{YUROV1991,DELFINO_1995,DELFINO1996327,DELFINO1996469,Delfino_2004}.
A relevant and complete process of the conformal bootstrap can be found in the appendix of
Ref.~\cite{xiao_cascade_2021}.
With another relevant field energy density (corresponding to $\sigma^{x}$ in the lattice model),
we can use $E_{8}$ form factor theory and bootstrap approach to
determine the SDSFs $S^{xx}(\omega)$, $S^{zz}(\omega)$.
And $S^{yy}(\omega)$ can be obtained via an exact relation
$S^{yy}(\omega) = S^{zz}(\omega) \omega^2/(2gJ)^2 $ \cite{jianda_E8_2014}.
Fig.~\ref{fig:E8_DSF_analytic} shows
the analytical SDSF with transfer momentum $q=0$
(corresponding to the Brillouin zone center in lattice),
which includes contributions from single- and multi-$E_{8}$ particle channels
up to total energy $5m_{1}$.

\begin{figure}[h]
\centering
\includegraphics[width=0.7\textwidth]{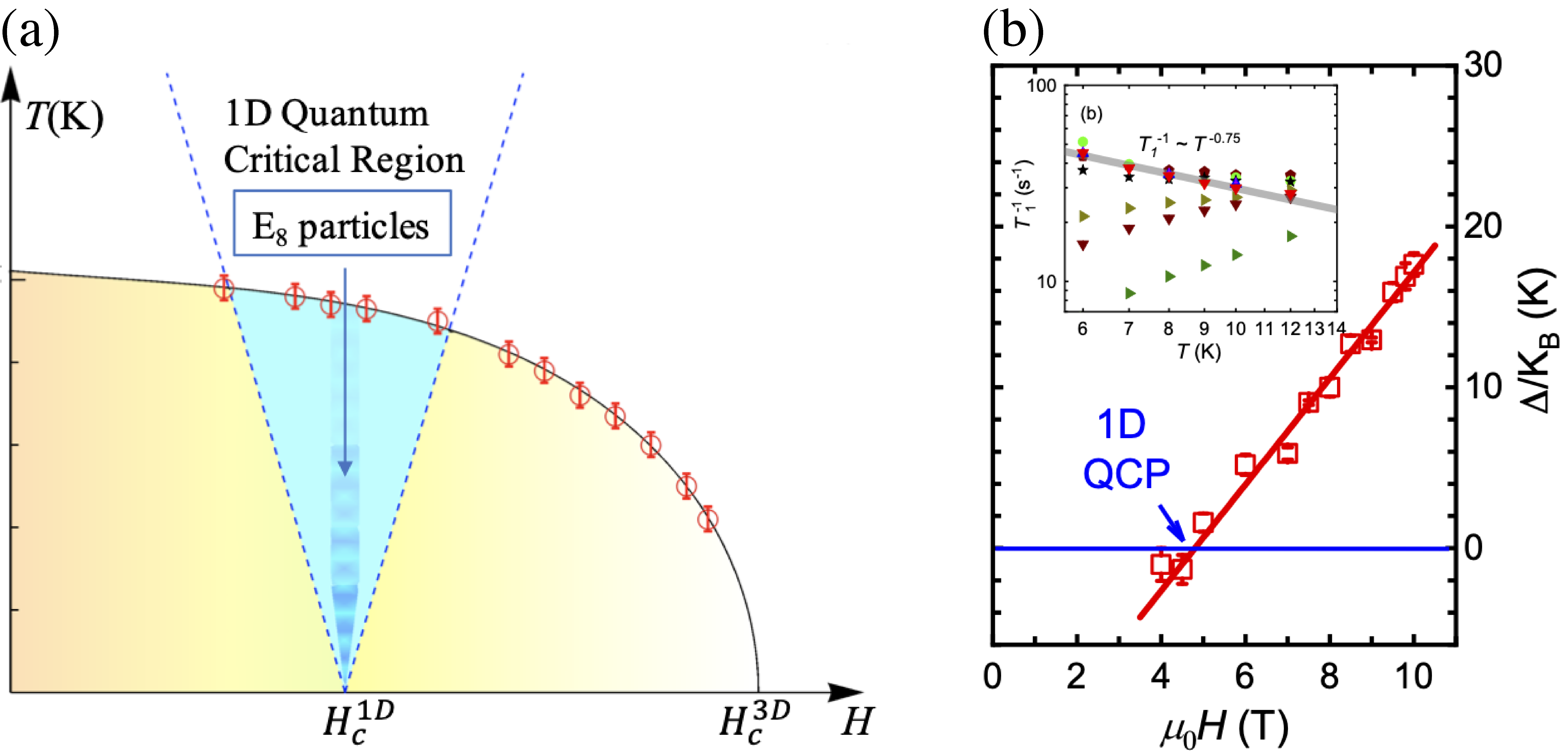}
\caption{
(a) Phase diagram of the BCVO with a transverse field along [010] direction;
(b) the quantum criticality of TFIC universality verified by NMR experiment.
From~\cite{zou_e_8_2021}.
}
\label{fig:E8_phase_NMR_TFIC}
\end{figure}

\begin{figure}[h]
\centering
\includegraphics[width=0.7\textwidth]{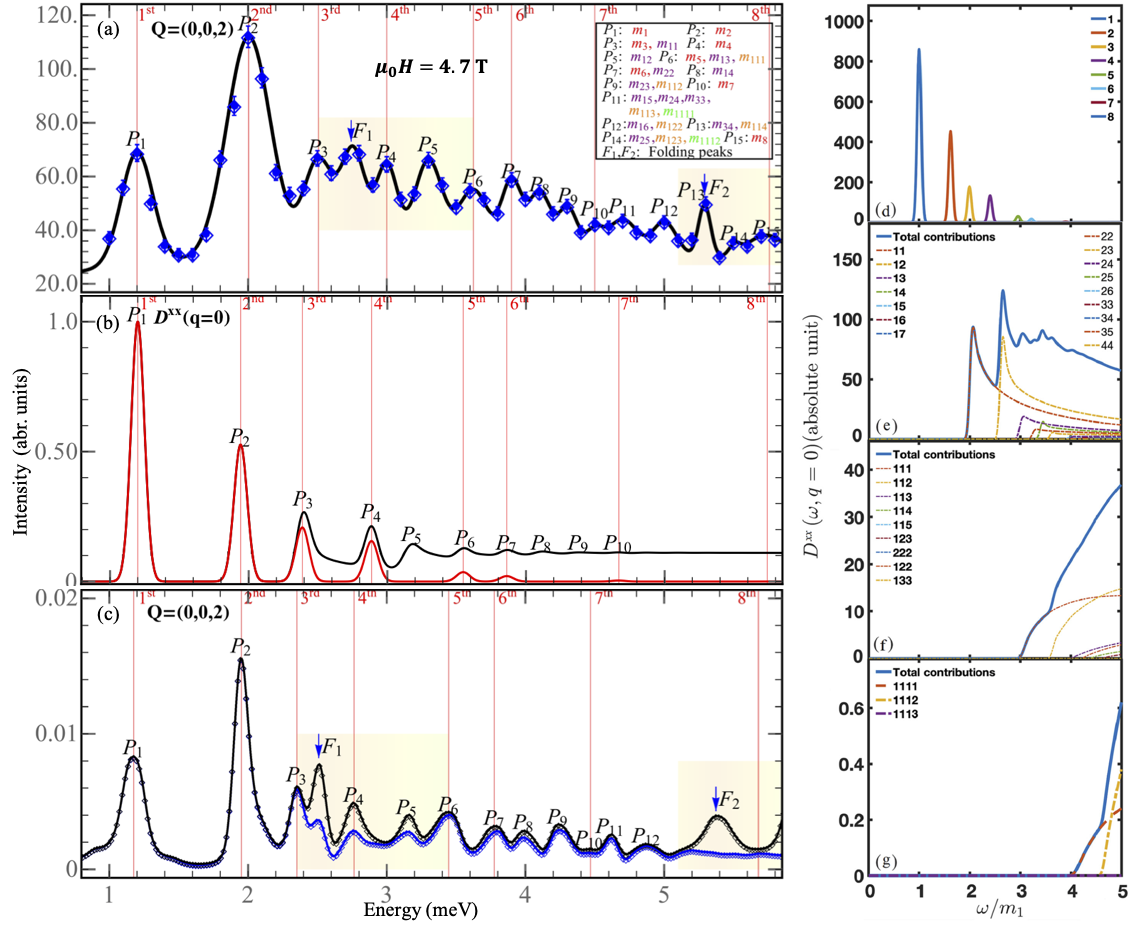}
\caption{The $E_{8}$ spectrum. (a) INS data, (b) analytical results, and (c) numerical results, where $0.1m_{1}$ energy resolution
broadening is applied to the latter two results
for comparison with INS data. (d), (e), (f) and (g) reveal the analytical
contributions of the spectrum from different $E_{8}$ channels. From~\cite{zou_e_8_2021}.}
\label{fig:E8_DSF}
\end{figure}

The existence of the TFIC universality in the BCVO and SCVO materials brings
in a hope to realize the $E_{8}$ model.
In order to realize the model in the AFM materials,
the perturbed longitudinal field needs to be a stagger field
along $z$ direction (the Ising spin direction) from site to site along the chain.
It is extremely difficult to directly apply such a field for condensed-matter experiments.
Different from SCVO where the 1D QCP is always outside its 3D ordering dome,
for BCVO when applying a field along [100] (or [010]) direction the field strength for
the corresponding 1D QCP is at $H_C^{1D} \approx 4.7$T $< H_C^{3D} \approx 10 $T, deep inside the BCVO's
3D ordering AFM dome [Fig.~\ref{fig:E8_phase_NMR_TFIC}].
Inside the 3D ordering phase, due to the weak interchain coupling,
we can conveniently apply chain mean-field theory to effectively
describe the corresponding physics,
whose effective Hamiltonian now becomes
\be
\small
\begin{split}
H&=H_{XXZ}
- H_x \sum_{i=1}^N\left\{ S_{i}^{x}
+h_y(-1)^{i}S_{i}^{y}
+h_zS^{z}_{i}
\cos\left[{\pi(2i-1)}/{4}
\right]
\right\}
-H'\sum_{i}(-1)^{i}S_{i}^{z},
\label{eq:XXZ_screw1}
\end{split}
\ee
where $J=5.8$~meV in $H_{XXZ}$ and the staggered perturbation $-H'\sum_{i}(-1)^{i}S_{i}^{z}$ with $H'=0.018J$ comes from the chain mean-field
of the interchain coupling, which can not be neglected in the 3D AFM ordering region.
The effective staggered field along $z$ direction
now can serve as the key perturbation field to realize the quantum $E_{8}$ model.
Near $H_C^{\rm 1D}$, the INS is carried out
to probe the magnetic excitations in the BCVO,
and the long-desired $E_8$ spectrum consists of all the eight single-$E_8$ particle peaks
and multi-particle continuum are observed for the first time~\cite{zou_e_8_2021}, as shown in Fig.~\ref{fig:E8_DSF}.
This result is fully consistent with the analytical calculation and the iTEBD calculation ~\cite{zou_e_8_2021}.

It is worth discussing a little more about how the comparison is completed, since it is a  nontrivial task to compare continuous results  (of field theory) with discrete results (of lattice calculations and experiments).
In the field theory the mass of lightest $E_8$ 
particle $m_1$ is related to the longitudinal field
as $m_1 = C \lambda^{8/15}$, which plays the role of
IR cutoff energy scale. When down to lattice 
we have a simple estimation $m_1^{lattice} = C' H'^{8/15} \approx 1.20$ meV where
$C' \approx 0.91C$ \cite{CASELLE2000667}.
The estimated $m_1^{lattice}$ is close to the 
real experimental observation,
therefore, we can safely pick up the first peak in the experimental data as the corresponding $m_1$ 
in the $E_8$ field theory.
Now we can make a detailed comparison of 
the analytical and experimental (numerical) zero-transfer-momentum DSF by using $m_1$ and the first peak of 
experimental (and numerical data) to re-scale their DSF spectra, respectively.
The comparison as shown in Fig.~\ref{fig:E8_DSF}, 
shows excellent agreement between discrete and analytical results,
which allows us to confidently claim that the 
$E_8$ physics is indeed observed in real experiments.

In the comparison, we should also note that when taking the space continuum limit,
$J$ for the nearest spin interaction in the lattice model indeed should serve as a UV energy cutoff for the applicability of the corresponding quantum field theory.
However, since $J$ is the energy scale for local spin interaction in the lattice, then, after taking the space continuum limit,
it actually appears as a cutoff scale for the energy of local or short-range fluctuations in the quantum field theory.
In the experiment (Fig.~\ref{fig:E8_DSF}),
the observed excitations at the zone center are coherent excitations involving large amounts of spins at long-range scale (zone center implies zero transfer momentum thus corresponding to long-range scale).
The excitation energy at such long-range scale cannot be simply considered
as the same energy for the local or short-range spin interaction.
Although it is highly nontrivial to accurately determine the energy
of local fluctuation (here it is the strength of nearest spin interaction)
from a many-body wavefunction,
it should be safe to claim that even the energy cost to get
a many-body coherent excitation at long-range scale is close to $J$,
on average the corresponding energy of local fluctuation should be still far smaller than $J$.
This should provide an underlying physical reason for the excellent comparison up to one $J$ as
shown in Fig.~\ref{fig:E8_DSF}.
Actually, even the energy cost is beyond $J$ for the long-range-scale
(corresponding to small (transfer) momentum) coherent excitation,
the field-theory prediction can still agree well with the lattice result,
as demonstrated in Ref.~\cite{Keselman_Dynamical_2020} which carefully
compares field theory results based on Bosonization with lattice calculation of the Heisenberg chain,
and in Ref.~\cite{Wybo_Quantum_2022} which carefully compares sine-Gordon field theory results
with lattice calculation of ladder XXZ chain.
Therefore, when considering the applicability of quantum field theory 
to the corresponding lattice model, it is not simply about energy scale (or time scale), 
the momentum scale (or length scale) also plays a crucial role.

\section{Conclusion}
\label{sec:conclusion}

This review details rich physics in the 1D spin-1/2 XXZ model with
the presence of various external fields and their experimental realization.
Because the experimental realization is
highly relevant to the quasi-1D AFM materials,
we first provide a systematic and pedagogical analysis for
setting up the effective Hamiltonian for the BCVO and SCVO,
which can also be applied to other
Co-based materials with cautions on possible
changes of symmetry for the local environment.
Then, we review recent theoretical progress on a variety of
magnetic excitaions of the 1D spin-1/2 XXZ model,
such as spinon (without magnetization),
(anti)psinon and string excitations (with finite magnetization),
and their experimental observations.
With the presence of transverse field, we further discuss
how the TFIC universality and the
exotic $E_8$ physics
(with additional small longitudinal field)
emerge near the corresponding 1D QCP, as well as their material realizations.

The 1D spin-1/2 XXZ model with various external fields contains
rich magnetic physics, whose exotic excitations
from low to high energy
are now successfully realized in real materials.
Those concrete progresses
can further inspire and bridge many research fields
such as quantum statistical field theory,
cold atom, and AdS/CFT etc..
Moreover, it is worth to further explore physical properties of those
exotic magnetic excitations, which will be an important step
toward practical control and application of those exotic magnetic
excitations.

\vskip 1.0 truecm

\noindent {\bf Acknowledgements}\\
The work is support by the Innovation Program for Quantum Science and Technology
No. 2021ZD0301900, the Natural Science Foundation of Shanghai with grant No. 20ZR1428400, and Shanghai Pujiang Program with grant No. 20PJ1408100. J.W. acknowledges additional support from a Shanghai talent program.

\appendix
\section{Cubic field potential in octahedron}
\label{app:Oct_V}

We consider the potential $\mathcal{V}(x,y,z)$ provided
by the ions with charge $q$ at the corners of an octahedron;
i.e. at $(\pm a,0,0)$, $(0,\pm a,0)$ and $(0,0,\pm a)$ [Fig.~\ref{fig:Co_octa} in main text].
Then it reads,
\be
\begin{split}
\mathcal{V}(x,y,z)
&=
V_x+V_y+V_z
\\&=
\sum_{\alpha=x,y,z}
\left[
\frac{q}{\left( r^2+a^2-2a\alpha \right)^{1/2}}
+
\frac{q}{\left( r^2+a^2+2a\alpha \right)^{1/2}}
\right]
\\&=
\frac{q}{\sqrt{A}}
\left[(1+X)^{-1/2}+(1-X)^{-1/2}
\right.\\&\quad\left.
(1+Y)^{-1/2}+(1-Y)^{-1/2}
+(1+Z)^{-1/2}+(1-Z)^{-1/2}\right]
\end{split}
\label{eq:V_xyz}
\ee
where $r^2=x^2+y^2+z^2$ is an arbitrary spatial point near the origin and
\be
\begin{split}
&A=(r^2+a^2),\quad B=\frac{r^2}{a^2},\quad
% \\&
X=\frac{2ax}{A}=\frac{2x}{a}\frac{1}{1+B},
\quad
Y=\frac{2ay}{A},
\quad
Z=\frac{2az}{A}.
\end{split}
\ee
Next, considering $a>r$, we expand Eq.~\eqref{eq:V_xyz} up to terms with sixth order of $r/a$.
Note that,
\be
(1+X)^{-1/2}+(1-X)^{-1/2}
=2+\frac{3}{4}X^2
+\frac{35}{64}X^4+
\frac{231}{512}X^6
+\mathcal{O}\left(\frac{r^8}{a^8}\right),
\ee
Eq.~\eqref{eq:V_xyz} becomes,
\be
\begin{split}
\mathcal{V}(x,y,z)&=
\frac{q}{\sqrt A}
\left[
6+\frac{3}{4}(X^2+Y^2+Z^2)
+\frac{35}{64}(X^4+Y^4+Z^4)+
\frac{231}{512}(X^6+Y^6+Z^6)
\right]
\\&\quad
\frac{q}{a}
\left\{
6-3\frac{r^2}{a^2}
+\frac{9}{4}\frac{r^4}{a^4}
-\frac{15}{8}
\frac{r^6}{a^6}
+\frac{r^2}{a^2}
\left(
3-\frac{15}{2}\frac{r^2}{a^2}+\frac{105}{8}\frac{r^4}{a^4}
\right)
\right.\\ &\quad \left.
+\frac{x^4+y^4+z^4}{a^4}
\left( \frac{35}{4}-\frac{315}{8}\frac{r^2}{a^2} \right)
+\frac{693}{24}\frac{(x^6+y^6+z^6)}{a^6}
\right\}
+\mathcal{O}\left(\frac{r^8}{a^8}\right).
\end{split}
\ee
\\
After some manipulations, we can obtain the desired expression,
\be
\begin{split}
\mathcal{V}(x,y,z)&=
\frac{6q}{a}
+\frac{35q}{4a^5}
\left[
(x^4+y^4+z^4)-\frac{3}{5}r^4
\right]
+\frac{-21q}{2a^7}
\\&\quad
\left[
(x^6+y^6+z^6)
+\frac{15}{4}(x^2y^4+x^2z^4+y^2x^4+y^2z^4+z^2x^4+z^2y^4)
-\frac{15}{14}r^6
\right]
\end{split}
\ee
where $\frac{35q}{4a^5}$ and $\frac{-21q}{2a^7}$ are prefactors $A_4$ in Eq.~\eqref{eq:V4} and $A_6$ Eq.~\eqref{eq:V6}, respectively.

For an alternative way, we can express the potential $\mathcal{V}(r,\theta,\phi)$ in terms of spherical harmonics, or tesseral harmonics.
The tesseral harmonics are defined as,
\be
\begin{split}
Z_{n0}&=Y_n^0,
\\
Z_{nm}^+&=\frac{1}{\sqrt 2}
[Y_n^{-m}+(-1)^mY_n^m],\quad \quad m>0
\\
Z_{nm}^-&=\frac{i}{\sqrt 2}
[Y_n^{-m}-(-1)^mY_n^m],\quad \quad m>0
\end{split}
\ee
which are always real and can be easily converted to cartesian coordinate functions.
Then, the potential $\mathcal{V}(r,\theta,\phi)$ reads,
\be
\mathcal{V}(r,\theta,\phi)=
\sum_n^{\infty}\sum_{\alpha}
r^n\gamma_{n\alpha}
Z_{n\alpha}(\theta,\phi)
\ee
and
\be
\gamma_{n\alpha}=
\sum_{j=1}^k
q_j
\frac{4\pi}{(2n+1)}
\frac{Z_{n\alpha}}{R_{j}^{(n+1)}},
\label{eq:gamma}
\ee
where the sum runs over all the neighboring anions, $k=6$ in our octahedron case.
Based on the analysis in the main text, only $\gamma_{00}$, $\gamma_{40}$, $\gamma_{60}$, $\gamma_{44}^+$ and $\gamma_{64}^+$ have non-vanishing contributions.
And using Eq.~\eqref{eq:gamma}, they can be obtained
\be
\begin{split}
&\gamma_{00}=\frac{12\sqrt{\pi}}{a}q
,\quad
\gamma_{40}=\frac{7\sqrt{\pi}}{3}\frac{q}{a^5}
,\quad
\gamma_{60}=\frac{3\sqrt{\pi}}{2\sqrt{13}}\frac{q}{a^7}
,\\&
\gamma_{44}^+=\frac{\sqrt{35\pi}}{3}\frac{q}{a^5}
,\quad
\gamma_{64}^+=-\frac{3\sqrt{7\pi}}{2\sqrt{13}}\frac{q}{a^7}.
\end{split}
\ee
Then, if we neglect the $Y_0^0$, the potential $\mathcal{V}(r,\theta,\phi)$ becomes
the desired form,
\be
\begin{split}
\mathcal{V}(r,\theta,\phi)&=
\frac{7\sqrt{\pi}}{3}\frac{q}{a^5}
\left[
Y_4^0(\theta,\varphi)
+\sqrt{\frac{5}{14}}
\left(
Y_4^4(\theta,\varphi)
+
Y_4^{-4}(\theta,\varphi)
\right)
\right]r^4
\\&+
\frac{3\sqrt{\pi}}{2\sqrt{13}}\frac{q}{a^7}
\left[
Y_6^0(\theta,\varphi)
-\sqrt{\frac{7}{2}}
\left(
Y_6^4(\theta,\varphi)
+Y_6^{-4}(\theta,\varphi)
\right)
\right]r^6
\end{split}
\ee
where $\frac{7\sqrt{\pi}}{3}\frac{q}{a^5}$ and
$\frac{3\sqrt{\pi}}{2\sqrt{13}}\frac{q}{a^7}$ are
prefactors $D_4$ in Eq.~\eqref{eq:V4} and $D_6$ Eq.~\eqref{eq:V6}, respectively.

\bibliographystyle{iopart-num}
\bibliography{Refs}

\end{document}